\documentclass[12pt]{iopart}
\expandafter\let\csname equation*\endcsname=\relax
\expandafter\let\csname endequation*\endcsname=\relax
\usepackage{amsmath, amssymb, color, graphicx, stmaryrd, wasysym}
\definecolor{linkcolor}{rgb}{0,0,0.6} 
\usepackage[colorlinks=true,
	pdfstartview = FitV,
	linkcolor    = linkcolor,
	citecolor    = linkcolor,
	urlcolor     = linkcolor,
	hyperindex   = true,
	hyperfigures = false]{hyperref}

\newcommand{\dd}{\text{d}}
\newcommand{\ee}{\text{e}}
\newcommand{\ii}{\text{i}}

\begin{document}

\title[Dissipation controls transport and phase transitions in active fluids]{Dissipation controls transport and phase transitions in active fluids: Mobility, diffusion and biased ensembles}

\author{\'Etienne Fodor}
\address{DAMTP, Centre for Mathematical Sciences, University of Cambridge, Wilberforce Road, Cambridge CB3 0WA, UK}

\author{Takahiro Nemoto}
\address{Philippe Meyer Institute for Theoretical Physics, Physics Department, \'Ecole Normale Sup\'erieure \& PSL Research University, 24 rue Lhomond, 75231 Paris Cedex 05, France}

\author{Suriyanarayanan Vaikuntanathan}
\address{James Franck Institute, University of Chicago, Chicago, IL 60637}
\address{Department of Chemistry, University of Chicago, Chicago, IL 60637}

\begin{abstract}
	Active fluids operate by constantly dissipating energy at the particle level to perform a directed motion, yielding dynamics and phases without any equilibrium equivalent. The emerging behaviors have been studied extensively, yet deciphering how local energy fluxes control the collective phenomena is still largely an open challenge. We provide generic relations between the activity-induced dissipation and the transport properties of an internal tracer. By exploiting a mapping between active fluctuations and disordered driving, our results reveal how the local dissipation, at the basis of self-propulsion, constrains internal transport by reducing the mobility and the diffusion of particles. Then, we employ techniques of large deviations to investigate how interactions are affected when varying dissipation. This leads us to shed light on a microscopic mechanism to promote clustering at low dissipation, and we also show the existence of collective motion at high dissipation. Overall, these results illustrate how tuning dissipation provides an alternative route to phase transitions in active fluids.
\end{abstract}

\maketitle


\section{Introduction}

Active matter comprises systems where energy is injected at the level of individual constituents~\cite{Marchetti2013, Bechinger2016, Stark2016, Marchetti2018}. Canonical examples at the micro-scale are given by swimming organisms which can be either biological or synthetic, such as bacteria~\cite{Libchaber2000, Elgeti2015} or Janus colloids in a fuel bath~\cite{Bechinger2013, Palacci2013}. These swimmers are able to convert a source of energy present in the environment, must often in chemical form, to perform a directed motion. Thus, the dynamics is  driven by a sustained flux of energy between the system components and their environment, so that active matter operates far from any equilibrium state~\cite{Speck2016, Nardini2016, Mandal2017, Seifert2018, Shankar2018, Bo2019}.

A striking property of active systems is their ability to spontaneously form clusters of particles even though interactions are purely repulsive, up to a complete separation between dense and dilute phases at large scales~\cite{Tailleur2008, Cates2015}. To rationalize such a phase transition, several works have relied on hydrodynamic theories. Some systematic coarse-graining procedures allow one to maintain the connection with the microscopic dynamics~\cite{Tailleur2013, Speck2013, Speck2014, Solon2015}. Alternatively, continuum descriptions which capture the phase separation are postulated based on symmetry arguments~\cite{Stenhammar2013, Wittkowski2014, Tiribocchi2015, Stenhammar2016, Nardini2017}. Despite the success of these approaches, the role of energy fluxes at the basis of directed motion is either completely discarded when mapping the dynamics into equilibrium~\cite{Tailleur2013, Farage2015, Brader2017}, or simply overlooked by using equilibrium-like tools, such as pressure and chemical potential, to describe these nonequilibrium transitions~\cite{Solon2018, Solon2018b}. Hence, determining how local energy flows control phase transitions at large scales remains an open challenge.

A phenomenological mechanism to describe cluster formation is based on the fact that collisions between active particles slow down the dynamics~\cite{Bechinger2016, Cates2015}. The reduction of transport coefficients by activity is then often regarded as a precursor of cluster formation. While several works have strived to predict how internal transport is affected by activity~\cite{Takatori2017, Stenhammar2017, Brady2017, Reichhardt2018, Saintillan2018, Tailleur2017}, a recent study has put forward an explicit connection between diffusion and dissipation in a mixture of active and passive particles~\cite{Suri2019}. Moreover, for generic driven systems, it has been shown recently that the diffusion coefficient is generically bounded by dissipation~\cite{Barato2015, Pietzonka2016a, Gingrich2016, Dechant2017, Garrahan2018}. Yet, this thermodynamic uncertainty relation (TUR) explicitly involves observable currents, thus being mostly useful for systems exhibiting directed transport. It remains to determine how TUR can also inform the transport in active systems in the absence of macroscopic particle current.

In this paper, we explore how local energy dissipation controls the transport of particles and the spontaneous formation of clusters in active fluids. In Sec.~\ref{sec:model}, we introduce the {\it collisional efficiency} to quantify cluster formation in terms of energy transfers between the particles and their environment. To predict how this efficiency behaves in terms of microscopic details, we build on the mapping of active particles into driven particles  proposed in~\cite{Suri2019}. It consists in describing the random self-propulsion as a disordered drive, which allows us to coarse-grain the active dynamics using methods of driven fluids, as detailed in Sec.~\ref{sec:map}. We then put forward in Sec.~\ref{sec:energy} generic relations between efficiency and internal transport: the efficiency equals a reduced mobility at high activity, and it also bounds the diffusion coefficient in such a regime. Finally, we demonstrate in Sec.~\ref{sec:bias} that modulating the efficiency with a specific control parameter amounts to changing the self-propulsion statistics. This result relies on methods of large deviations where atypical dynamics are promoted to select a desired level of efficiency~\cite{Lecomte2007, Touchette2009, Jack2010}, leading to promote either cluster formation or collective motion. Altogether, our results illustrate how efficiency relates to internal transport and quantifies the emergence of phase transitions, thus supporting that changing efficiency is a generic route to control the properties of active fluids.


\section{Energy transfers in active fluids}\label{sec:model}

We consider a set of self-propelled particles immersed in a solvent at temperature $T$ and interacting through a potential $U$. Provided that inertia is negligible, the dynamics is given by an overdamped Langevin equation:
\begin{equation}\label{eq:dyn}
	\dot{\bf r}_i = {\bf v}_i - \nabla_i U + {\boldsymbol\xi}_i ,
\end{equation}
where we have set the particle mobility to unity. The thermal fluctuations stemming from the solvent are accounted for by the zero-mean Gaussian white noise $ {\boldsymbol\xi}_i $ with correlations
\begin{equation}
	\langle \xi_{i\alpha}(t) \xi_{j\beta}(0) \rangle = 2 T \delta_{ij} \delta_{\alpha\beta} \delta(t) ,
\end{equation}
where the Latin and Greek indices respectively refer to particle labels and spatial components. Following recent works~\cite{Szamel2015, Maggi2015, Farage2015, Nardini2016}, we describe the self-propulsion force ${\bf v}_i$ as another zero-mean Gaussian noise, uncorrelated with ${\boldsymbol\xi}_i$, with correlations
\begin{equation}\label{eq:active_corr}
	\langle v_{i\alpha}(t) v_{j\beta}(0) \rangle = \frac{T_\text{\tiny A}}{\tau} \delta_{ij} \delta_{\alpha\beta} \ee^{-|t|/\tau} ,
\end{equation}
where $\tau$ is the persistence time, and $T_\text{\tiny A}$ is the energy scale of active fluctuations. The self-propulsion correlations~\eqref{eq:active_corr} correspond to the Ornstein-Uhlenbeck process
\begin{equation}\label{eq:v}
	\tau\dot{\bf v}_i=-{\bf v}_i+{\boldsymbol\eta}_i ,
\end{equation}
where ${\boldsymbol\eta}_i$ is a zero-mean Gaussian white noise, uncorrelated with ${\boldsymbol\xi}_i$, with correlations:
\begin{equation}
	\langle\eta_{i\alpha}(t)\eta_{j\beta}(0) \rangle = 2T_\text{\tiny A}\delta_{ij}\delta_{\alpha\beta}\delta(t) .
\end{equation}
Such a model has then been referred to as Active Ornstein-Uhlenbeck Particles (AOUPs). For a vanishing persistence, the active fluctuations cannot be distinguished from the thermal ones: $\langle v_{i\alpha}(t)v_{j\beta}(0)\rangle \underset{\tau\to0}{\longrightarrow} 2T_\text{\tiny A} \delta_{ij}\delta_{\alpha\beta}\delta(t)$, in which case the system amounts to a set of passive Brownian particles at temperature $T+T_\text{\tiny A}$. The deviation from this equilibrium regime is controlled by the ratio of persistence time $\tau$ to equilibrium relaxation time $a^2/T$, where $a$ is the particle diameter, defined as the P\'eclet number~\cite{Nardini2016}:
\begin{equation}
	\text{Pe} = \frac{\sqrt{T\tau}}{a} .
\end{equation}
Alternatively, setting directly $T_\text{\tiny A}=0$ corresponds the equilibrium regime at temperature $T$ for any value of $\tau$.

The amount of dissipated energy $\cal J$ is defined from purely mechanical arguments as the rate of work that the particles exert on the solvent~\cite{Sekimoto1998, Seifert2012, Kanazawa2014, Ahmed2016}:
\begin{equation}\label{eq:diss}
	{\cal J} = \langle \dot{\bf r}_i \cdot (\dot{\bf r}_i - {\boldsymbol\xi}_i ) \rangle .
\end{equation}
Repeated indices are implicitly summed, and the product rule $\cdot$ is defined within Stratonovitch convention, as for the rest of the paper. In driven systems, such a definition coincides with the rate of entropy production, which can be deduced from the ratio of forward to backward path probability weights~\cite{Sekimoto1998, Seifert2012, Lebowitz1999}. In active systems, different definitions of entropy production have been proposed~\cite{Speck2016, Nardini2016, Mandal2017, Seifert2018, Shankar2018, Bo2019}, the connection with the dissipation $\cal J$ is only maintained for some of them.

Substituting the dynamics~\eqref{eq:dyn} in the dissipation definition~\eqref{eq:diss}, the dissipation can be written as	${\cal J} = (1/\tau)\langle{\bf r}_i\cdot{\bf v}_i\rangle$, where we have used $\langle \dot{\bf r}_i \cdot \nabla_i U \rangle = \dd \langle U\rangle / \dd t= 0$ and $\langle\dot{\bf r}_i\cdot{\bf v}_i\rangle - (1/\tau)\langle{\bf r}_i\cdot{\bf v}_i\rangle = \dd\langle{\bf r}_i\cdot{\bf v}_i\rangle/\dd t = 0$ in steady state. Thus, $\cal J$ is proportional to the swim pressure $P_{\rm s} = (\rho_0/d) \langle {\bf r}_0 \cdot {\bf v}_0 \rangle$, where ${\bf v}_0$ and ${\bf r}_0$ respectively refer to the self-propulsion force and the position of a given particle, $\rho_0$ is the particle density and $d$ is the spatial dimension~\cite{Marchetti2014, Brady2014, Solon2015b}. The swim pressure is known to be the main contribution to the total pressure at low density, where it can be written for short-range repulsive interactions as $P_{\rm s} = \rho_0 T_\text{\tiny A} (1 - \rho_0/\rho_\text{c})$, where $\rho_\text{c}$ is the density near close packing. This provides a direct relation between dissipation and density:	${\cal J}=(d N T_\text{\tiny A}/\tau) (1 - \rho_0/\rho_\text{c})$ when $\rho_0<\rho_\text{c}$.

\begin{figure}
	\centering
	\includegraphics[width=.5\columnwidth]{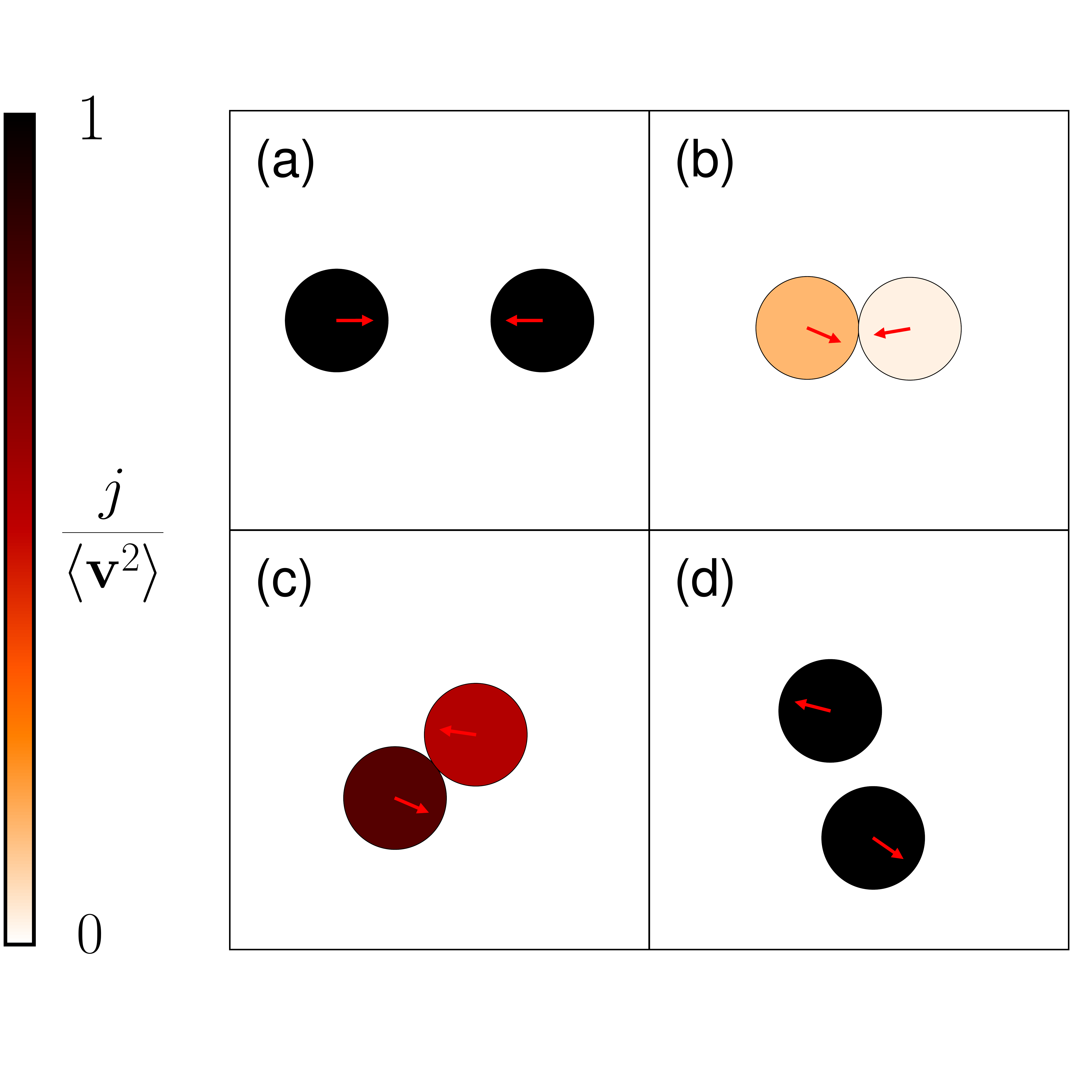}
	\caption{\label{fig:diss}
	Snapshots of colliding active particles at four successive times from (a) to (d). The red arrows denote the direction of self-propulsion. The color code refers to the instantaneous dissipation $j = {\bf v} \cdot ({\bf v}-\nabla U) $ scaled by the free-motion average value	$\langle{\bf v}^2\rangle$. It is maximum when the particles are separated, as shown in (a,d), and its value gets reduced during the collision, as depicted in (b,c). Movie in~\cite{Supplemental}.
	}
\end{figure}

For arbitrary density and interaction, the dissipation can still be separated into distinct contributions stemming from free motion and from interactions:
\begin{equation}
	{\cal J} = \frac{d N T_\text{\tiny A}}{\tau} - \langle{\bf v}_i\cdot\nabla_iU\rangle ,
\end{equation}
where we have used that ${\bf v}_i$ and ${\boldsymbol\xi}_i$ are uncorrelated, and $\langle{\bf v}_i^2\rangle = d N T_\text{\tiny A}/\tau$. The free motion term $d N T_\text{\tiny A}/\tau$ is the maximum amount of dissipation. Collisions reduce the dissipation by slowing-down particles, as shown in Fig.~\ref{fig:diss}, which points at a natural connection between collision-induced cluster formation and reduced dissipation. To quantify the propensity of collisions to stabilize clusters, we then introduce the {\it collisional efficiency}, denoted $\cal E$, which compares the contribution to dissipation from interactions and from free motion :
\begin{equation}\label{eq:eff_def}
	{\cal E} = \frac{\tau}{d N T_\text{\tiny A}} \langle{\bf v}_i\cdot\nabla_iU\rangle .
\end{equation}
It takes values between $0$ and $1$, respectively in the absence of interactions and when all particles are in stable clusters. Besides, it is proportional to the ``rate of work'' introduced recently in nonequilibrium liquids as a measure of the drive-induced dynamical and structural changes~\cite{Suri2019, Junco2018a}. With this definition, the efficiency is small when most of the dissipation is due to free motion and the fluid only features a small number of clusters, whereas the efficiency is large when particles dissipate less power in the solvent by forming large clusters. Hence, the efficiency quantifies the ability of collisions among particles to yield clustering by slowing-down the dynamics.

\section{Coarse-graining active fluids}\label{sec:map}

To study the relation between particle-based efficiency and collective behavior, we use a coarse-grained description of the dynamics. It generally consists in a set of hydrodynamic equations for the density $\rho({\bf r},t) = \sum_i \delta[{\bf r}-{\bf r}_i(t)]$ and the polarization ${\boldsymbol{\cal P}}({\bf r},t) = \sum_i {\bf v}_i(t) \delta[{\bf r}-{\bf r}_i(t)]$. The conservation equation for density can be explicitly derived from the microscopic dynamics~\eqref{eq:dyn} as
\begin{equation}\label{eq:dyn_dens_init}
	\partial_t \rho = \nabla \cdot \Big[ \rho\nabla U-{\boldsymbol{\cal P}} + T \nabla \rho + \sqrt{2\rho T} {\boldsymbol\Lambda} \Big] ,
\end{equation}
where the fluctuating term ${\boldsymbol\Lambda}$ is a zero-mean Gaussian noise with correlations
\begin{equation}
	\langle \Lambda_\alpha({\bf r},t) \Lambda_\beta({\bf r}',t') \rangle = \delta_{\alpha\beta} \delta({\bf r}-{\bf r}') \delta(t-t') .
\end{equation}
In the absence of aligning interactions, instabilities can be detected solely based on the time evolution of density. To obtain a closed dynamics for density only, the challenge is then to express the polarization in terms of the density. This commonly relies on combining adiabatic treatments with appropriate closures, valid only in asymptotic regimes. While such treatments have shown successful to capture the onset of phase transitions~\cite{Tailleur2013, Speck2013, Speck2014, Solon2015}, they are generally restricted to low activity, thus limiting their ability to describe properly fluctuations in the homogeneous state.

Our goal is then to bridge the gap between the hydrodynamics of fluids driven by external forces, obtained from a systematic derivation without reference to any closure~\cite{Kawasaki1994, Dean1996}, and the hydrodynamics of active fluids. To this aim, we rely on the fact that the active fluctuations of the self-propulsion can be mapped into a deterministic drive with disordered amplitude. Following a recent work~\cite{Suri2019}, we introduce a set of driving forces ${\bf f}_i$ describing periodic orbits in the reference frame of each particle as
\begin{equation}\label{eq:drive}
	{\bf f}_i(t) = \sqrt{\frac{T_\text{\tiny A}}{n\tau}} \sum_{a=1}^n \big[ {\bf A}_{ai} \cos(\omega_at) +  {\bf B}_{ai} \sin(\omega_at) \big] .
\end{equation}
The frequencies $\{\omega_i\}$ are identical for all particles, and the amplitudes $\{{\bf A}_{ai}, {\bf B}_{ai}\}$ are taken as uncorrelated zero-mean Gaussian variables with unit variance for each particle:
\begin{equation}\label{eq:amp}
	\langle A_{ai\alpha} A_{bj\beta} \rangle_\text{d} = \delta_{ab}\delta_{ij}\delta_{\alpha\beta} = \langle B_{ai\alpha} B_{bj\beta} \rangle_\text{d} ,
\end{equation}
where $\langle\cdot\rangle_\text{d}$ denotes an average over the disorder. Then, the disordered drive ${\bf f}_i$ is a zero-mean Gaussian process with correlations given for a large number of oscillators ($n\gg1$) as
\begin{equation}\label{eq:drive_corr}
	\langle f_{i\alpha} (t) f_{j\beta}(0) \rangle_\text{d} \underset{n\gg1}{=} \delta_{ij} \delta_{\alpha\beta} \frac{T_\text{\tiny A}}{\tau} \int \phi(\omega) \ee^{\ii\omega|t|} \frac{\dd\omega}{2\pi} .
\end{equation}
where $\phi$ is the density of driving frequencies. In this limit, ${\bf f}_i$ then maps into a random noise with spectrum given by $\phi$. Specifically, by choosing the following density
\begin{equation}\label{eq:g}
	\phi(\omega) = \frac{2\tau}{1 + (\omega\tau)^2} ,
\end{equation}
the drive statistics~\eqref{eq:drive_corr} reproduces the exponential decay of active fluctuations~\eqref{eq:active_corr}. When the typical frequency of the system response, set by the inverse relaxation time $1/\tau_\text{\tiny R}$, is below the drive frequency $1/\tau$, the disordered drive cannot be distinguished from a white noise: this is the small P\'eclet regime ($\text{Pe}\ll1$). In contrast, the response of the system probes the non-flat part of the driving density when the cut-off frequency $1/\tau$ is below the relaxation frequency $1/\tau_\text{\tiny R}$: this is the high P\'eclet regime ($\text{Pe}\gg1$), where the drive statistics is now effectively colored.

Within such a mapping, we can now derive hydrodynamic equations using the standard techniques of driven fluids~\cite{Kawasaki1994, Dean1996}. The first step consists in identifying the source of noise in the disordered dynamics~\eqref{eq:dyn}, where now ${\bf v}_i$ is replaced by ${\bf f}_i$. This is straightforward in the asymptotic P\'eclet regimes: (i)~for $\text{Pe}\ll1$, the drive ${\bf f}_i$ is akin to a white noise, whose fluctuations can be absorbed into a modified solvent temperature $T+T_\text{\tiny A}$, and (ii)~for $\text{Pe}\gg1$, the drive ${\bf f}_i$ is seemingly a deterministic force with constant amplitude and direction. Whether the drive should be regarded as random or deterministic is ambiguous in between these regimes, so that the coarse-graining cannot be employed {\it a priori}.

To treat simultaneously both the large and small P\'eclet regimes, we define modified thermal noise $\tilde{\boldsymbol\xi}_i$ and driving force $\tilde{\bf f}_i$, where the former is the only source of fluctuations in the dynamics and the latter is a purely deterministic drive. This is achieved by enforcing (i) $\langle \tilde\xi_{i\alpha}(t) \tilde\xi_{j\beta}(0) \rangle = 2 (T+T_\text{\tiny A}) \delta_{ij} \delta_{\alpha\beta} \delta(t)$ and $\tilde f_{i\alpha} = 0$ when $\text{Pe}\ll1$, and (ii) $\langle \tilde\xi_{i\alpha}(t) \tilde\xi_{j\beta}(0) \rangle = 2 T\delta_{ij} \delta_{\alpha\beta} \delta(t)$ and $\tilde{\bf f}_i = {\bf f}_i$ when $\text{Pe}\gg1$. In short, the correlations of the tilted processes can be written in terms of the dimensionless parameter $\sigma$ as
\begin{equation}\label{eq:drive_corr_g}
	\begin{aligned}
		\langle \tilde\xi_{i\alpha}(t) \tilde\xi_{j\beta}(0) \rangle &= 2 (T+\sigma T_\text{\tiny A}) \delta_{ij} \delta_{\alpha\beta} \delta(t) ,
		\\
		\langle \tilde f_{i\alpha} (t) \tilde f_{j\beta}(0) \rangle_\text{d} &= \langle f_{i\alpha} (t) f_{j\beta}(0) \rangle_\text{d} - 2 \sigma T_\text{\tiny A} \delta_{ij} \delta_{\alpha\beta} \delta(t) ,
	\end{aligned}
\end{equation}
where $\sigma=1$ when $\text{Pe}\ll1$, and $\sigma=0$ when $\text{Pe}\gg1$. This provides an explicit connection between the statistics of $\{{\bf f}_i,{\boldsymbol\xi}_i\}$ and that of $\{\tilde{\bf f}_i,\tilde{\boldsymbol\xi}_i\}$ in the asymptotic P\'eclet regimes.

It follows that the dynamics~\eqref{eq:dyn} can be written equivalently, with now clearly separated random and deterministic contributions, as
\begin{equation}\label{eq:dyn_map_g}
	\dot{\bf r}_i = \tilde{\bf f}_i - \nabla_i U + \tilde{\boldsymbol\xi}_i .
\end{equation}
Note that the noise term $\tilde{\boldsymbol \xi}_i$ contains contributions from both active and thermal fluctuations at small $\text{Pe}$. Thus, the thermodynamic interpretation of~\eqref{eq:dyn_map_g} is deliberately lost, so that it should only be regarded as a convenient effective dynamics. Using the methods in~\cite{Kawasaki1994, Dean1996}, the dynamics of the density field at fixed disorder then reads
\begin{equation}\label{eq:dyn_dens}
	\partial_t \rho = \nabla \cdot \Big[ \rho\nabla U-{\bf F} + (T+\sigma T_\text{\tiny A}) \nabla \rho + \sqrt{2\rho(T+\sigma T_\text{\tiny A})} {\boldsymbol\Lambda} \Big] ,
\end{equation}
where ${\bf F} ({\bf r},t) = \sum_i \tilde{\bf f}_i(t) \delta[{\bf r}-{\bf r}_i(t)]$ is the driving field. In short, we have turned the hydrodynamic equation~\eqref{eq:dyn_dens_init}, written for particles subject to thermal and active fluctuations with respective temperatures $T$ and $T_\text{\tiny A}$, into the closed hydrodynamics~\eqref{eq:dyn_dens} of thermal-like particles at temperature $T+\sigma T_\text {\tiny A}$ subject to an external random drive $\bf F$. This allows us to obtain the dynamics of the density fluctuations at fixed disorder without any closure on the statistics of the polarization $\boldsymbol{\cal P}$.

From this effective dynamics, we now characterize the fluctuations of a given active particle which acts as probe of the bath formed by surrounding particles. To this aim, we introduce the reduced fields $\bar\rho({\bf r},t) = \sum_{i\neq0} \delta[{\bf r}-{\bf r}_i(t)]$ and $\bar{\bf F}({\bf r},t) = \sum_{i\neq0} \tilde{\bf f}_i(t) \delta[{\bf r}-{\bf r}_i(t)]$ which characterize the dynamics of particles other than tracer, where ${\bf r}_0$ is the tracer position. Considering pair-wise interactions of the form $U(\{{\bf r}_i\}) = \sum_{i<j} V({\bf r}_i-{\bf r}_j)$, the tracer position evolves according to
\begin{equation}\label{eq:dyn_tracer_bath}
	\dot{\bf r}_0 = \tilde{\bf f}_0 - \int \bar\rho({\bf r},t) \nabla_0 V({\bf r}-{\bf r}_0) \dd {\bf r} + \tilde{\boldsymbol\xi}_0 ,
\end{equation}
where the density dynamics follows readily from~\eqref{eq:dyn_dens} as
\begin{equation}\label{eq:dyn_dens_bath}
	\begin{aligned}
		\partial_t \bar\rho({\bf r},t) &= \nabla \cdot \bigg[ \bar\rho({\bf r},t) \nabla \bigg(\int \bar\rho({\bf r}',t) V({\bf r}-{\bf r}') \dd{\bf r}' + V({\bf r}-{\bf r}_0) \bigg) + (T+\sigma T_\text{\tiny A})\nabla\bar\rho({\bf r},t) \bigg]
		\\
		&\quad + \nabla \cdot \Big[ \sqrt{2\bar\rho({\bf r},t)(T+\sigma T_\text{\tiny A})} {\boldsymbol\Lambda}({\bf r},t) - \bar{\bf F}({\bf r},t) \Big].
	\end{aligned}
\end{equation}
Following~\cite{Demery2011, Demery2014}, the collective modes $\delta\rho_{\bf k}(t) = \int[ \bar\rho({\bf r},t) - \rho_0 ] \ee^{-\ii{\bf k}\cdot{\bf r}} \dd\bf r$, can be linearized around the homogeneous profile $\rho_0$ for weak interactions as
\begin{equation}\label{eq:modes}
	\delta\rho_{\bf k}(t) = \int \dd s {\cal G}_{\bf k}(t-s) \Big\{ - {\bf k}^2 \rho_0 V_{\bf k} \ee^{-\ii{\bf k}\cdot{\bf r}_0(s)} + \ii{\bf k}\cdot \Big[ \sqrt{2\rho_0(T+\sigma T_\text{\tiny A})} {\boldsymbol\Lambda}_{\bf k}(s) - \bar{\bf F}_{\bf k}(s) \Big] \Big\} ,
\end{equation}
where ${\cal G}_{\bf k}(t) = \ee^{-{\bf k}^2(T+\sigma T_\text{\tiny A}+\rho_0 V_{\bf k})t} \Theta(t) $.

In practice, such a linearization only requires the tracer-bath interactions to be weak, without any further specifications on bath-bath interactions, since density fluctuations are generically Gaussian in the homogeneous phase~\cite{Chandler1993, Fily2012}. To highlight this, we introduce a small dimensionless parameter $h\ll1$ which scales the bath-tracer potential as $V({\bf r})=h\bar V({\bf r})$. Substituting~\eqref{eq:modes} in~\eqref{eq:dyn_tracer_bath}, the tracer dynamics then follows in a closed form as
\begin{equation}\label{eq:dyn_tracer}
	\begin{aligned}
		\dot{\bf r}_0&(t) - \rho_0 h^2 \int\dd s\int_{\bf k} {\cal G}_{\bf k}(t-s) \ii{\bf k} (|{\bf k}|\bar V_{\bf k})^2 \ee^{ \ii{\bf k}\cdot[{\bf r}_0(t)-{\bf r}_0(s)] }
		\\
		&= \tilde{\bf f}_0(t) - h \int\dd s\int_{\bf k} {\cal G}_{\bf k}(t-s) {\bf k} \bar V_{\bf k} \ee^{\ii{\bf k}\cdot{\bf r}_0(t)} \big[ {\bf k}\cdot\bar{\bf F}_{\bf k}(s) \big] + {\boldsymbol\Gamma}[{\bf r}_0(t),t] + \tilde{\boldsymbol\xi}_0(t) ,
	\end{aligned}
\end{equation}
where $\int_{\bf k}=\int\dd{\bf k}/(2\pi)^d$, and ${\boldsymbol\Gamma}$ is a zero-mean Gaussian noise with correlations
\begin{equation}
	\langle \Gamma_\alpha[{\bf r}_0(t),t] \Gamma_\beta[{\bf r}_0(0),0] \rangle = \delta_{\alpha\beta} \rho_0h^2(T+\sigma T_\text{\tiny A}) \int_{\bf k} \frac{{\cal G}_{\bf k}(|t|) (|{\bf k}|\bar V_{\bf k})^2 \ee^{\ii{\bf k}\cdot[{\bf r}_0(t)-{\bf r}_0(0)]}}{T + \sigma T_\text{\tiny A} + \rho_0 V_{\bf k}} .
\end{equation}
Thus, the effect of the bath on the tracer dynamics can be separated into three contributions: (i)~a damping term reflecting the effect of the tracer on the surrounding particles, which in turn is opposed to the tracer displacement, (ii)~a forcing term which embodies how driving the surrounding particles affect the tracer, and (iii)~a multiplicative noise term. Note that the tracer dynamics~\eqref{eq:dyn_tracer} is given at fixed disorder without any prior assumption on the values of the P\'eclet number, since our derivation only relies on a perturbation in terms of $h$.

As a result, following the procedure in~\cite{Demery2011, Demery2014}, we have reduced the many-body description of the system into an effective dynamics for the tracer only. This leads to introduce memory effects containing explicit details about the interactions with surrounding particles. Even though the tracer statistics extracted from~\eqref{eq:dyn_tracer} should only be accurate for weak interactions {\it a priori}, it has been shown that qualitative features remain relevant even beyond such a regime~\cite{Demery2015, Martin2018, Demery2019}. Note that our derivation does not rely on any response theory at variance with recent works~\cite{Steffenoni2016, Maes2017, England2018}. Moreover, we consider here that both the tracer and its surrounding particles are active, in contrast with~\cite{Suri2019} which only considered a dilute fraction of active tracers in a bath of passive particles, and the tracer statistics can now be derived for an arbitrary strength of the active force.


\section{Efficiency and transport}\label{sec:energy}

\begin{figure}
	\centering
	\includegraphics[width=.7\columnwidth]{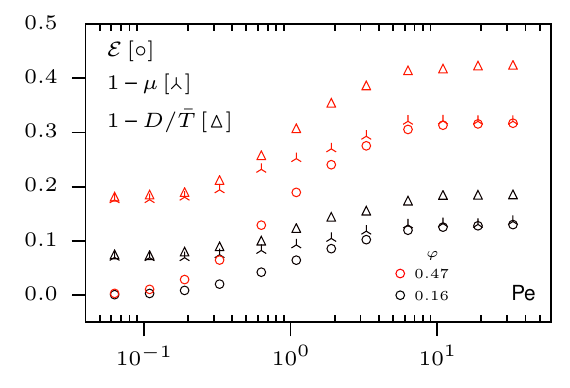}
	\caption{\label{fig:eff}
		Collisional efficiency $\cal E$, reduced mobility $1-\mu$ and reduced diffusion coefficient $1-D/\bar T$, where $\bar T=T+T_\text{\tiny A}$, as functions of the P\'eclet number $\text{Pe}$, for a fluid of active particles with dynamics~\eqref{eq:dyn}. Numerical measurements reveal a generic behavior where $1-\mu$ evolves between two plateau values: (i) it equals $1-D/\bar T$ for $\text{Pe}\ll1$, as expected from the Stokes-Einstein relation, and (ii) it equals $\cal E$ for $\text{Pe}\gg1$.
		Parameter values: $L=10$, $a=1$, $V_\text{\tiny M}=50T$, $T=T_\text{\tiny A}$, $f_\text{\tiny P}=5\times10^{-2}$.
	}
\end{figure}

To illustrate how energy transfers constrain fluctuations, we now turn to deriving generic relations between efficiency and transport coefficients. In what follows, we consider particles interacting {\it via} short-range soft repulsion with a pair-wise potential of the form $V({\bf r}) = V_\text{\tiny M} (1-r/a)^2 \Theta(a-r)$, where $\Theta$ denotes the Heaviside step function. We perform numerical simulations of the original dynamics~\eqref{eq:dyn} in two dimensions using periodic boundary conditions in a box of size $L$. We focus on regimes where the system does not undergo a complete phase separation between a single cluster and a dilute phase, but rather consists of separated clusters constantly forming and disintegrating. The efficiency $\cal E$ as a function of the P\'eclet number vanishes at low activity ($\text{Pe}\ll1$), namely when only a negligible amount of clusters are present, as shown in Fig.~\ref{fig:eff}. It increases with $\text{Pe}$, as a signature of the increasing number of clusters, and it saturates at high activity ($\text{Pe}\gg1$). When decreasing the packing fraction $\varphi=\rho_0\pi a^2/4$, the plateau value at large $\text{Pe}$ gets reduced since less clusters are formed.

To capture this behavior analytically, we derive an expression for the efficiency $\cal E$ in terms of microscopic details. Our derivation consist in first evaluating $\cal E$ at fixed disorder, using the effective tracer dynamics~\eqref{eq:dyn_tracer}, and then averaging over disorder to account for active fluctuations. Within the mapping in Sec.~\ref{sec:map}, any observable depending explicitly on the self-propulsion ${\bf v}_i$ can be computed at fixed disorder by simply replacing ${\bf v}_i$ with the corresponding driving force ${\bf f}_i$. The efficiency~\eqref{eq:eff_def} can then written as ${\cal E} = 1 - (\tau/dNT_\text{\tiny A}) \langle\langle\dot{\bf r}_i \rangle\cdot{\bf f}_i \rangle_\text{d}$, where $\langle\cdot\rangle$ denotes here an average at fixed disorder. As detailed in~\ref{app:tracer}, we determine the average tracer velocity $\langle\dot{\bf r}_i \rangle$ perturbatively for weak interactions, and the efficiency follows directly by averaging over disorder, see~\eqref{eq:eff}. This result does not rely on any perturbation in terms of the P\'eclet number, yet evaluating the noise correlations~\eqref{eq:drive_corr_g}, as controlled by $\sigma$, is only explicit for asymptotic P\'eclet regimes. In such regimes, the leading order of efficiency gets simplified as
\begin{equation}\label{eq:eff_as}
	\begin{aligned}
		{\cal E} &\underset{\text{Pe}\ll1}{=} \frac{\tau\rho_0}{d} \int_{\bf k} \frac{(|{\bf k}|V_{\bf k})^2}{T+T_\text{\tiny A}+\rho_0V_{\bf k}} ,
		\\
		{\cal E} &\underset{\text{Pe}\gg1}{=} \frac{\rho_0}{d} \int_{\bf k} \frac{V_{\bf k}^2}{(T+\rho_0V_{\bf k})(2T+\rho_0V_{\bf k})} ,
	\end{aligned}
\end{equation}
where we have absorbed the small parameter $h$ back into the bare definition of the bath-tracer interaction potential $V_{\bf k}=h\bar V_{\bf k}$, and we have neglected the contribution of orders higher than $h^2$. Importantly, our predictions are valid for arbitrary $T_\text{\tiny A}$, in contrast with~\cite{Suri2019} where derivations are performed perturbatively in terms of the driving amplitude. Besides, all particles are active in our case, whereas~\cite{Suri2019} only considered a small fraction in a bath of passive particles.

\begin{figure}
	\centering
	\includegraphics[width=.7\columnwidth]{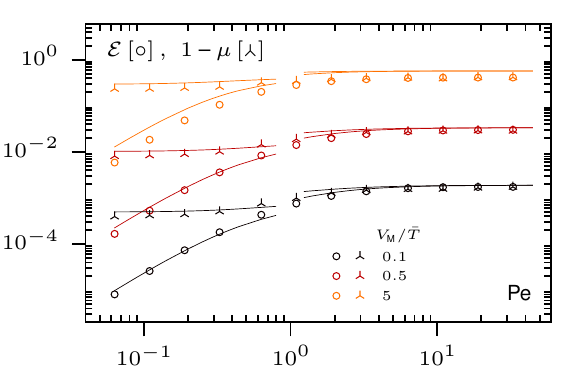}
	\caption{\label{fig:mob}
		Collisional efficiency $\cal E$ and reduced mobility $1-\mu$ as functions of the P\'eclet number $\text{Pe}$. Numerical results are reported for increasing values of the scaled interaction strength $V_\text{\tiny M}/\bar T$, where $\bar T = T+T_\text{\tiny A}$. The solid lines refer to the corresponding analytic predictions in~\eqref{eq:eff} and~\eqref{eq:mob}, taking $\sigma=\{1,0\}$ respectively for ${\rm Pe}<1$ and ${\rm Pe}>1$.
		Parameter values: $L=10$, $a=1$, $\rho_0=1$, $T=T_\text{\tiny A}$, $f_\text{\tiny P}=5\times10^{-2}$.
	}
\end{figure}

To probe the range of validity of our results, we compare them with simulations of the original dynamics~\eqref{eq:dyn} for increasing interaction strength $V_\text{\tiny M}$, as shown in Fig.~\ref{fig:mob}. By taking $T=T_\text{\tiny A}$ at all values of the P\'eclet number, we enforce that the dynamics remains away from any phase separation. At weak interactions, we observe a quantitative agreement with numerics for a wide range of $\text{Pe}$ values. Interestingly, the predictions at small and large $\text{Pe}$, corresponding respectively to $\sigma=\{1,0\}$ in~\eqref{eq:eff}, actually reproduce the numerics even close to ${\rm Pe}=1$ where one would expect them to fail a priori. This supports that our analytic approach, based on separating the averages over white noise and over disorder, captures correctly the tracer dynamics not only in the asymptotic P\'eclet regimes, but also for intermediate P\'eclet values. Deviations from numerics appear at higher values of $V_\text{\tiny M}$, yet we still observe the linear scaling and the saturation of $\cal E$ predicted by~\eqref{eq:eff_as} at small and large $\text{Pe}$, respectively.

To relate efficiency with internal transport, we consider the statistics of  tracer displacement. The linear mobility $\mu$ quantifies how a constant perturbation force $f_\text{\tiny P}\hat{\bf e}$ affects the average tracer velocity:
\begin{equation}\label{eq:mob_def}
	\mu = \underset{t\to\infty}{\lim} \frac{\langle\dot{\bf r}_0(t)\rangle\cdot\hat{\bf e}}{f_\text{\tiny P}} \bigg|_{f_\text{\tiny P}=0} .
\end{equation}
The diffusion coefficient $D$ characterizes the spontaneous tracer excursion in terms of the position variance as
\begin{equation}
	D = \underset{t\to\infty}{\lim} \frac{1}{2dt} \Big\langle \big[ \langle{\bf r}_0(t)\rangle - {\bf r}_0(t) \big]^2 \Big\rangle .
\end{equation}
In the absence of interactions, the tracer mobility is the solvent mobility, set to unity, and the diffusion coefficient equals the equilibrium temperature $T+T_\text{\tiny A}$. The effect of interactions is to resist the tracer displacement, thus reducing the transport coefficients. In what follows, we quantify how active fluctuations reduce the mobility and the diffusion coefficient.

We measure the reduction of mobility and of scaled diffusion coefficient, given respectively by $1-\mu$ and $1-D/(T+T_\text{\tiny A})$, as functions of the P\'eclet number by performing numerical simulations of the original dynamics~\eqref{eq:dyn}. We observe three regimes shown in Fig.~\ref{fig:eff}: (i)~both measurements coincide at small $\text{Pe}$ in agreement with the Stokes-Einstein relation for equilibrium fluids, (ii)~they increase monotonically with $\text{Pe}$ showing that more and more collisions hinder the tracer displacement, and (iii)~they saturate at large $\text{Pe}$ with distinct plateau values. The plateau values increase with the packing fraction $\varphi$, since the number of clusters resisting tracer displacement also increases. Interestingly, the reduced mobility $1-\mu$ converges to the efficiency $\cal E$ at large $\text{Pe}$. This can be rationalized within the mapping in Sec.~\ref{sec:map}. At fixed disorder, the scaled dissipation ${\cal J}/{\bf f}_j^2 = [\langle\dot{\bf r}_i\rangle\cdot{\bf f}_i] / {\bf f}_j^2$ can be regarded as the mobility measured with respect to ${\bf f}_i$. In the regime $\text{Pe}\gg1$, the drive is seemingly deterministic with small amplitude $ | {\bf f}_i | = \sqrt{T_\text{\tiny A}/\tau} $. Then, the scaled dissipation is similar to the linear mobility in such a regime, so that efficiency and mobility reduction should indeed coincide asymptotically.

To predict how the mobility behaves in terms of microscopic details, we rely again on the effective tracer dynamics~\eqref{eq:dyn_tracer}. Following the same procedure as for the efficiency, we use a perturbative expansion at weak interactions, see~\eqref{eq:mob}. The mobility reduction follows in the asymptotic regimes as
\begin{equation}\label{eq:mob_as}
	\begin{aligned}
		1-\mu &\underset{\text{Pe}\ll1}{=} \frac{\rho_0}{d} \int_{\bf k} \frac{V_{\bf k}^2}{(T+T_\text{\tiny A}+\rho_0V_{\bf k})[2(T+T_\text{\tiny A})+\rho_0V_{\bf k}]} ,
		\\
		1-\mu &\underset{\text{Pe}\gg1}{=} \frac{\rho_0}{d} \int_{\bf k} \frac{V_{\bf k}^2}{(T+\rho_0V_{\bf k})(2T+\rho_0V_{\bf k})} .
	\end{aligned}
\end{equation}
At small $\text{Pe}$, we recover the result for an equilibrium system at temperature $T+T_\text{\tiny A}$~\cite{Demery2014}. The plateau value at large P\'eclet coincides with the one of efficiency in~\eqref{eq:eff_as}, as expected. We observe a quantitative agreement between our prediction and numerical simulations at weak interactions in a large range of P\'eclet values, see Fig.~\ref{fig:mob}. Though deviations appear for stronger interactions, the measurements of reduced mobility $1-\mu$ always coincide with that of efficiency $\cal E$ at high $\text{Pe}$, thus illustrating how energy transfers constrain the tracer displacement in this regime.

We now explore how to relate the diffusion coefficient to efficiency. Recent works have shown that, in nonequilibrium dynamics, the current fluctuations are bounded by the coarse-grained entropy production rate~\cite{Barato2015, Pietzonka2016a, Gingrich2016, Dechant2017, Garrahan2018}. In active fluids, there is no average current of particles provided that no asymmetric external potential is applied~\cite{Leonardo2010, Reichhardt2017, Pietzonka2019}. Yet, the bound on current fluctuations can still be formulated within the mapping of Sec.~\ref{sec:map}. The coarse-grained entropy production rate $\Sigma$ quantifies the breakdown of time reversal symmetry in the hydrodynamic description, given at fixed disorder by~\eqref{eq:dyn_dens}. Following~\cite{Lebowitz1999, Seifert2012}, it is defined in terms of the probability weights for the forward and backward dynamics, respectively denoted by $\mathbb{P}_\rho$ and $\mathbb{P}^\text{\tiny R}_\rho$, as
\begin{equation}\label{eq:sigma}
	\Sigma = \underset{t\to\infty}{\lim} \frac{1}{t} \ln \frac{\mathbb{P}_\rho}{\mathbb{P}^\text{\tiny R}_\rho} .
\end{equation}
The bound on the particle current ${\bf V}({\bf r}, t)=\sum_i\dot{\bf r}_i(t)\delta[{\bf r}-{\bf r}_i(t)]$ can then be written at fixed disorder as
\begin{equation}\label{eq:bound}
	\frac{1}{D} \bigg[\int\langle{\bf V}\rangle\dd{\bf r}\bigg]^2 < \Sigma .
\end{equation}
Besides, we show in~\ref{app:ent} that $\Sigma$ can be written explicitly in terms of the current $\bf V$ and of the driving field ${\bf F}({\bf r},t)=\sum_i\tilde{\bf f}_i(t)\delta[{\bf r}-{\bf r}_i(t)]$ as
\begin{equation}\label{eq:sigma_g}
	\Sigma = \frac{1}{T+\sigma T_\text{\tiny A}} \int \bigg\langle \frac{{\bf V}\cdot{\bf F}}{\rho} \bigg\rangle \dd{\bf r} .
\end{equation}
Substituting the explicit definitions of $\{\rho, {\bf F}, {\bf V}\}$ and integrating over space yields
\begin{equation}
	\Sigma = \frac{\langle\dot{\bf r}_i\rangle\cdot\tilde{\bf f}_i}{T+\sigma T_\text{\tiny A}} ,
\end{equation}
where we have used the discernibility condition $\delta[{\bf r}-{\bf r}_i(t)]\delta[{\bf r}-{\bf r}_j(t)]\simeq\delta_{ij}\delta[{\bf r}-{\bf r}_i(t)]\delta[{\bf r}-{\bf r}_j(t)]$ which enforces that distinct particles cannot have identical positions at the same instant.

The value of $\sigma$ is arbitrary within this formulation. To obtain explicit result in terms of measurable observables, we focus on the high $\text{Pe}$ regime where $\sigma=0$, so that $\tilde{\bf f}_i$ and ${\bf f}_i$ have the same statistics according to~\eqref{eq:drive_corr_g}. In this regime, the coarse-grained entropy $\Sigma$ averaged over disorder is then simply related to dissipation $\cal J$ as
\begin{equation}\label{eq:sigma_av}
	\langle\Sigma\rangle_\text{d} \underset{\text{Pe}\gg1}{=} \frac{\cal J}{T} = \frac{1}{T}\bigg[\frac{d N T_\text{\tiny A}}{\tau} - \langle{\bf f}_i\cdot\langle\nabla_i U\rangle\rangle_\text{d}\bigg] .
\end{equation}
Similarly, when averaging over disorder, the squared average current in~\eqref{eq:bound} can be related to efficiency as follows. Substituting the microscopic dynamics~\eqref{eq:dyn} in ${\bf V}({\bf r}, t)=\sum_i\dot{\bf r}_i(t)\delta[{\bf r}-{\bf r}_i(t)]$ and considering weak interactions, we get 
\begin{equation}\label{eq:current_av}
	\begin{aligned}
		\bigg\langle\bigg[\int\langle{\bf V}\rangle\dd{\bf r}\bigg]^2\bigg\rangle_\text{d} &= \langle\langle\dot{\bf r}_i\rangle^2\rangle_\text{d}
		\\
		&=\frac{d N T_\text{\tiny A}}{\tau} - 2 \langle{\bf f}_i\cdot\langle\nabla_i U\rangle\rangle_\text{d} + {\cal O}(h^3) ,
	\end{aligned}
\end{equation}
where we have replaced ${\bf v}_i$ by ${\bf f}_i$ within the mapping of Sec.~\ref{sec:map}, and we have used again the discernibility condition. Combining~\eqref{eq:sigma_av} and~\eqref{eq:current_av}, we then deduce
\begin{equation}
	\frac{1}{\langle\Sigma\rangle_\text{d}} \bigg\langle\bigg[\int\langle{\bf V}\rangle\dd{\bf r}\bigg]^2\bigg\rangle_\text{d} \underset{\text{Pe}\gg1}{=} T(1-{\cal E}) + {\cal O}(h^3) .
\end{equation}
Finally, the dissipation bound~\eqref{eq:bound} involving the current statistics of the current can now be written independently of the current as
\begin{equation}\label{eq:bound_ac}
	\frac{D}{T} \underset{\text{Pe}\gg1}{>} 1 - {\cal E} ,
\end{equation}
where we have neglected the contribution of orders higher than $h^2$. As a result, the bound on current fluctuations translates for active fluids into a bound between diffusion coefficient and efficiency at high P\'eclet number. This bound can also be formulated in terms of diffusion coefficient and mobility, given that the latter coincides with the reduced efficiency $1-\cal E$ in this regime. Note that the results in Fig.~\ref{fig:eff} show that, in practice, the high P\'eclet regime is already reached for $\text{Pe}\gtrsim1$.

\begin{figure}
	\centering
	\includegraphics[width=1\columnwidth]{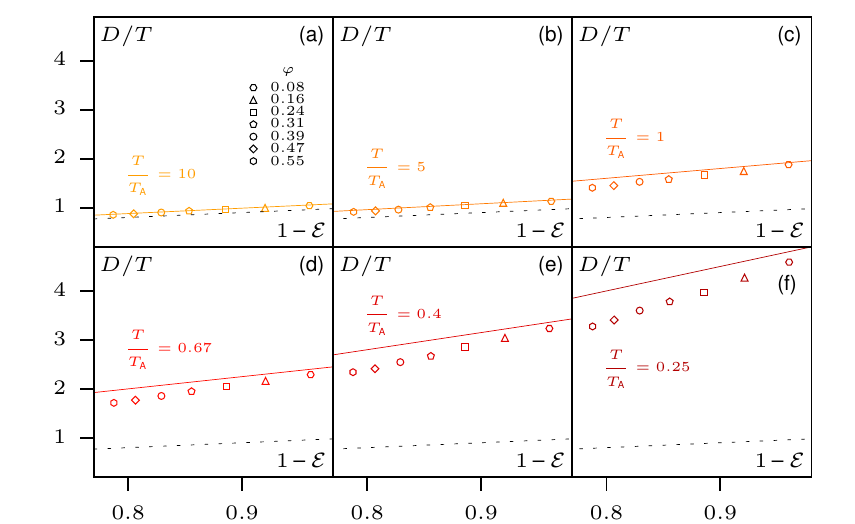}
	\caption{\label{fig:diff}
		Parametric plot of the scaled diffusion coefficient $D/T$ and the reduced collisional efficiency $1-{\cal E}$ in the regime of large P\'eclet number $\text{Pe}$, for different values of the packing fraction $\varphi$. Numerical results support the existence of a lower bound $D/T>1-{\cal E}$, in black dashed line, and an upper bound $(1+T/T_\text{\tiny A})(1-{\cal E})>D/T$, in colored solid line.
		Parameter values: $L=10$, $a=1$, $V_\text{\tiny M}=50$, $\tau=10^2$, $T=10$; (a)~$T_\text{\tiny A}=1$; (b)~$T_\text{\tiny A}=2$; (c)~$T_\text{\tiny A}=10$; (d)~$T_\text{\tiny A}=15$; (e)~$T_\text{\tiny A}=25$; (f)~$T_\text{\tiny A}=40$.
	}
\end{figure}

To quantify how far from the bound the dynamics operates, we measure the diffusion coefficient $D$ and the fluid efficiency $\cal E$ for different values of the packing fraction $\varphi$ and of the passive to active temperature ratio $T/T_\text{\tiny A}$. We report these measurements in a parametric plot of $D$ as a function of $1-\cal E$, equivalent to $D$ as a function of $\mu$ provided that the P\'eclet number is large enough, see Fig.~\ref{fig:diff}. Inspired by the numerical results in Fig.~\ref{fig:eff}, we also test the empirical bound $(1+T_\text{\tiny A}/T)(1-{\cal E}) < D/T$. When $T\gg T_\text{\tiny A}$, this bound combined with~\eqref{eq:bound_ac} enforces that the scaled diffusion $D/T$ coincides with the reduced efficiency $1-{\cal E}$, or equivalently with the mobility $\mu$: this is the Stokes-Einstein relation for an equilibrium fluid at temperature $T$. In practice, this relation is confirmed numerically for a large range of packing fractions at $T=10T_\text{\tiny A}$, see Fig.~\ref{fig:diff}(a).

When decreasing progressively $T/T_\text{\tiny A}$, the two bounds become increasingly separated as shown in Figs.~\ref{fig:diff}(b-f). The numerics are close to the empirical bound at low packing fraction, showing that the scaled diffusion $D/(T+T_\text{\tiny A})$ coincides with the mobility $\mu$ when interactions are negligible: this is the Stokes-Einstein relation for an equilibrium fluid at temperature $T+T_\text{\tiny A}$, as expected. Conversely, the distance from the bound~\eqref{eq:bound_ac} reduces with packing fraction, suggesting that the assumption of Gaussian density fluctuations, from which~\eqref{eq:bound} was originally derived, is essentially valid at high packing fraction. Overall, our numerical results confirm that the reduced efficiency $1-{\cal E}$ bounds the diffusion coefficient $D$, as another example of how energy transfers constrain  transport coefficients.


\section{Phase transitions in biased ensembles}\label{sec:bias}

We now turn to discuss how the efficiency controls phase transitions. To modulate efficiency with a specific control parameter, we rely on tools of large deviations which allow one to select an arbitrary value of efficiency by promoting atypical realisations of the dynamics~\cite{Lecomte2007, Touchette2009, Jack2010}. This has been used extensively in kinetically constrained dynamics and glassy models where promoting atypical currents leads to probe system configurations otherwise inaccessible~\cite{Merolle2005, Garrahan2007, Hedges2009, Speck2012}. In active systems, the connexion between clustering and rare fluctuations has been explored only recently, thus shedding light on phase transitions at constant activity and density~\cite{Suma2017, Whitelam2017, Whitelam2017b, Nemoto2018}.

In equilibrium, the energy is commonly changed by varying the temperature of the system, which controls the weight of {\it configurations} according to the Boltzmann distribution. By analogy, changing the efficiency out-of-equilibrium consists in biasing the path probability of the dynamics, thus constraining the particle {\it trajectories}. For convenience, we focus in what follows on regimes where thermal fluctuations are irrelevant ($T\ll T_\text{\tiny A}$), which amounts to neglecting ${\boldsymbol\xi}_i$ in~\eqref{eq:dyn}. The efficiency $\cal E$ can then be simplified as 
\begin{equation}\label{eq:eff_sim}
	{\cal E} \underset{T\ll T_\text{\tiny A}}{=} \frac{\tau}{dNT_\text{\tiny A}} \langle(\nabla_iU)^2\rangle ,
\end{equation}
where we have used $\langle\dot{\bf r}_i\cdot\nabla_iU\rangle=\dd\langle U\rangle/\dd t=0$. Introducing the time-extensive efficiency $\varepsilon$ as
\begin{equation}\label{eq:eff_sc}
	\varepsilon = \frac{\tau}{d N T_\text{\tiny A}} \int_0^t (\nabla_i U)^2 \dd s ,
\end{equation}
which is related to the bare efficiency as $\underset{t\to\infty}{\lim} (\varepsilon/t) = {\cal E}$, we then define the biased path probability $\mathbb{P}_\lambda[\{{\bf r}_i\}_0^t]$ in terms of the path probability of the original dynamics $\mathbb{P}[\{{\bf r}_i\}_0^t]$ as
\begin{equation}
	\mathbb{P}_\lambda [\{{\bf r}_i\}_0^t] \sim \mathbb{P} [\{{\bf r}_i\}_0^t] \,\ee^{ - \lambda \varepsilon } .
\end{equation}
The conjugate parameter $\lambda$ controls the efficiency analogously to how temperature controls the energy in equilibrium. Setting positive or negative values of $\lambda$ corresponds to enforcing respectively low or high efficiency, thus allowing us to probe how the system adapts for an arbitrary value of efficiency.

Given that the biased ensemble associated with $\mathbb{P}_\lambda$ does not conserve probability density, any connection with explicit dynamics is generally lost~\cite{Evans2004, Evans2005}. Yet, a systematic procedure allows one to potentially infer an auxiliary dynamics which effectively realizes the constraint on trajectories~\cite{Touchette2013, Jack2015, Chetrite2015}. The auxiliary dynamics considered previously are for exclusion processes~\cite{Popkov2010, Popkov2011, Limmer2018a, Cagnetta2019}, particle-based diffusive systems restricted to small noise regimes~\cite{Lecomte2019, Proesmans2019} and non-interacting cases in specific potentials~\cite{Majumdar2002, Touchette2016, Touchette2018}. Interestingly, recent works have also put forward explicit solutions in active systems for a mean-field dynamics~\cite{Limmer2018} and for a many-body dynamics with pair-wise forces~\cite{Suri2019}.

In practice, the Fokker-Planck operator of the auxiliary dynamics ${\cal L}_\text{aux}$ follows from the one of the original dynamics ${\cal L}$ according to a	 generalized Doob's transform~\cite{Touchette2013, Jack2015, Chetrite2015}. Provided that thermal fluctuations are neglected, the microscopic dynamics~(\ref{eq:dyn}-\ref{eq:v}) can be written in terms of positions ${\bf r}_i$ and velocities $\dot{\bf r}_i = {\bf p}_i$ as~\cite{Nardini2016}
\begin{equation}\label{eq:dyn_p}
	\tau\dot{\bf p}_i = -{\bf p}_i -(1+\tau{\bf p}_j\cdot\nabla_j) \nabla_iU + {\boldsymbol\eta}_i .
\end{equation}
The generalized Doob's transform is then given by
\begin{equation}\label{eq:aux}
	{\cal L}_\text{aux} = {\cal L} - \frac{2T_\text{\tiny A}}{\tau^2} \frac{\partial}{\partial{\bf p}_i} \cdot \frac{\partial\Phi}{\partial{\bf p}_i} ,
\end{equation}
where $\Phi(\{{\bf r}_i\},\{{\bf p}_i\};\lambda)$ satisfies the following eigenvalue problem
\begin{equation}\label{eq:eigen}
	\bigg[{\cal L}^\dagger - \frac{\lambda\tau}{dNT_\text{\tiny A}} (\nabla_iU)^2 \bigg]\,\ee^{\Phi} = \psi(\lambda)\,\ee^{\Phi} ,
\end{equation}
in terms of the operator ${\cal L}^\dagger$ adjoint to $\cal L$, and of the scaled cumulant generating function $\psi$ associated with $\varepsilon$. To propose an explicit solution for $\Phi$, we use a perturbative expansion at small P\'eclet number without any assumption regarding particle interactions, at variance with the approximations used in Secs.~\ref{sec:map} and~\ref{sec:energy}. We demonstrate in~\ref{app:bias} that the first order auxiliary dynamics is given by
\begin{equation}\label{eq:dyn_aux}
	\begin{aligned}
		\dot{\bf r}_i &= {\bf v}_i - \nabla_i U ,
		\\
		\tau\dot{\bf v}_i &= - {\bf v}_i + \kappa(\{{\bf r}_i\};\lambda) \big[\nabla_iU - {\bf v}_i\big] + {\boldsymbol\eta}_i ,
	\end{aligned}
\end{equation}
where $\kappa$ reads
\begin{equation}\label{eq:kappa_aux}
	\kappa(\{{\bf r}_j\};\lambda) = \bigg[ \frac{\lambda\tau}{dNT_\text{\tiny A}}(\nabla_kU)^2 + \psi(\lambda) \bigg] \,\ee^{\tau\dot{\bf r}_j^2/(2T_\text{\tiny A})} \, E_{1-dN/2} \bigg( \frac{\tau\dot{\bf r}_j^2}{2T_\text{\tiny A}} \bigg) ,
\end{equation}
in terms of the exponential integral function $E_n(z) = \int_1^\infty \dd t \ee^{-zt}/t^n$. As a result, changing efficiency by tuning the control parameter $\lambda$ amounts to introducing explicitly particle interactions in the self-propulsion dynamics, so that the self-propulsion is no longer an independent process for each particle. This clearly differs from a previous work where biasing with a dissipation-related observable led to renormalize particle interactions~\cite{Suri2019}.

Provided that collective rearrangements operate more slowly than the particle dynamics, there is generally a time scale separation between the fluctuations of $\kappa$ and that of ${\bf v}_i$. Then, $\kappa$ can be taken as approximately constant during a well-chosen time interval. When particles get arrested by collisions ($|\dot{\bf r}_i|=0$), the self-propulsion statistics is independent of interactions as in the unbiased case ($\lambda=0$), see~\eqref{eq:dyn_aux}, in which case one recovers the exponential decay of correlations in~\eqref{eq:active_corr}. When particles do not interact with neighbours ($|\nabla_iU|=0$), the self-propulsion correlations also decay exponentially, yet now with renormalized amplitude $T_\text{\tiny A}/(1+\kappa)^2$ and renormalized persistence $\tau/(1+\kappa)$. Then, we deduce that the amplitude and the persistence of self-propulsion are effectively different in dense regions of the system, where collisions are frequent, compared with dilute regions where particles are mostly moving freely.

\begin{figure}
	\centering
	\includegraphics[width=.49\columnwidth]{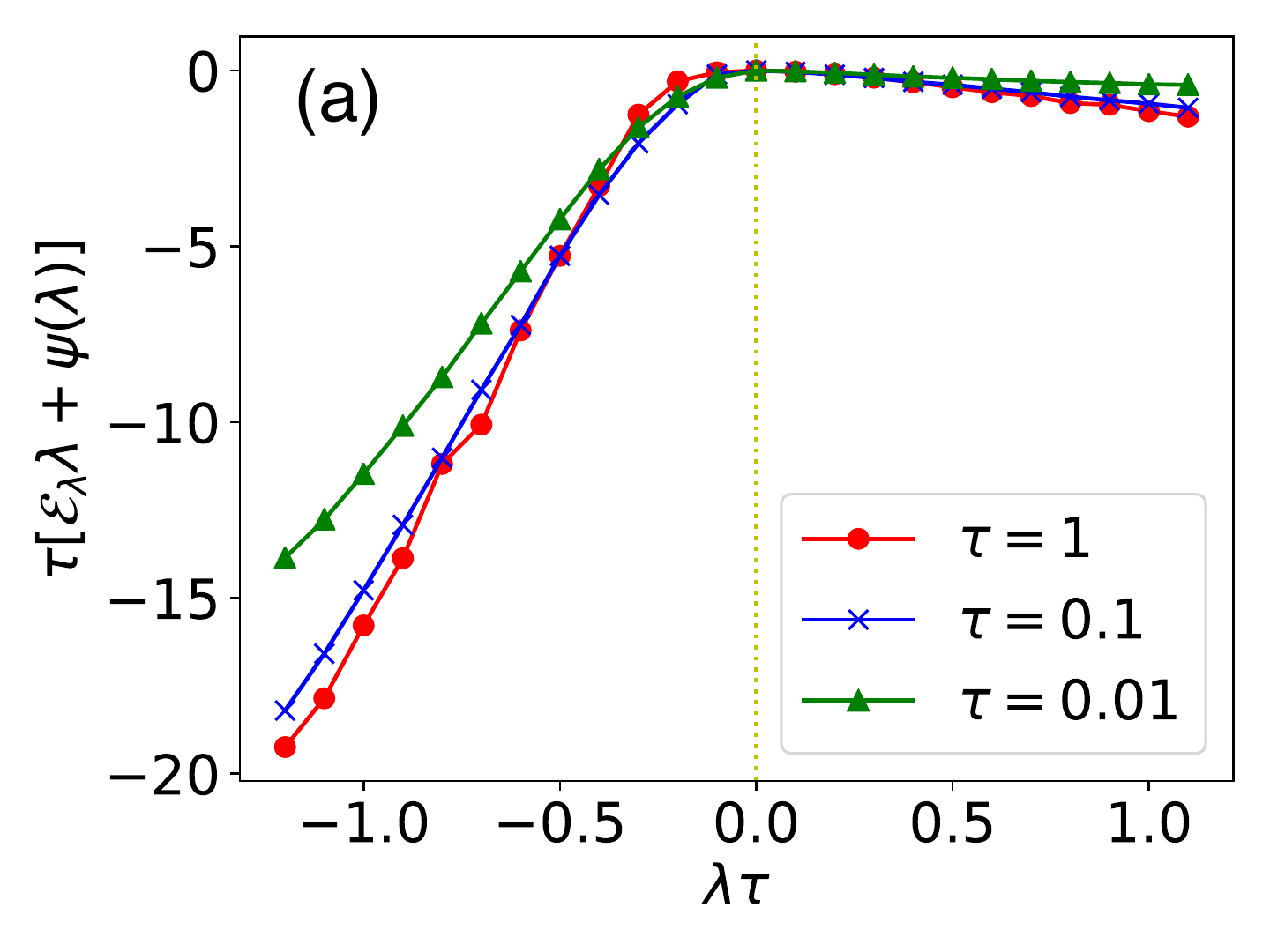}
	\includegraphics[width=.49\columnwidth]{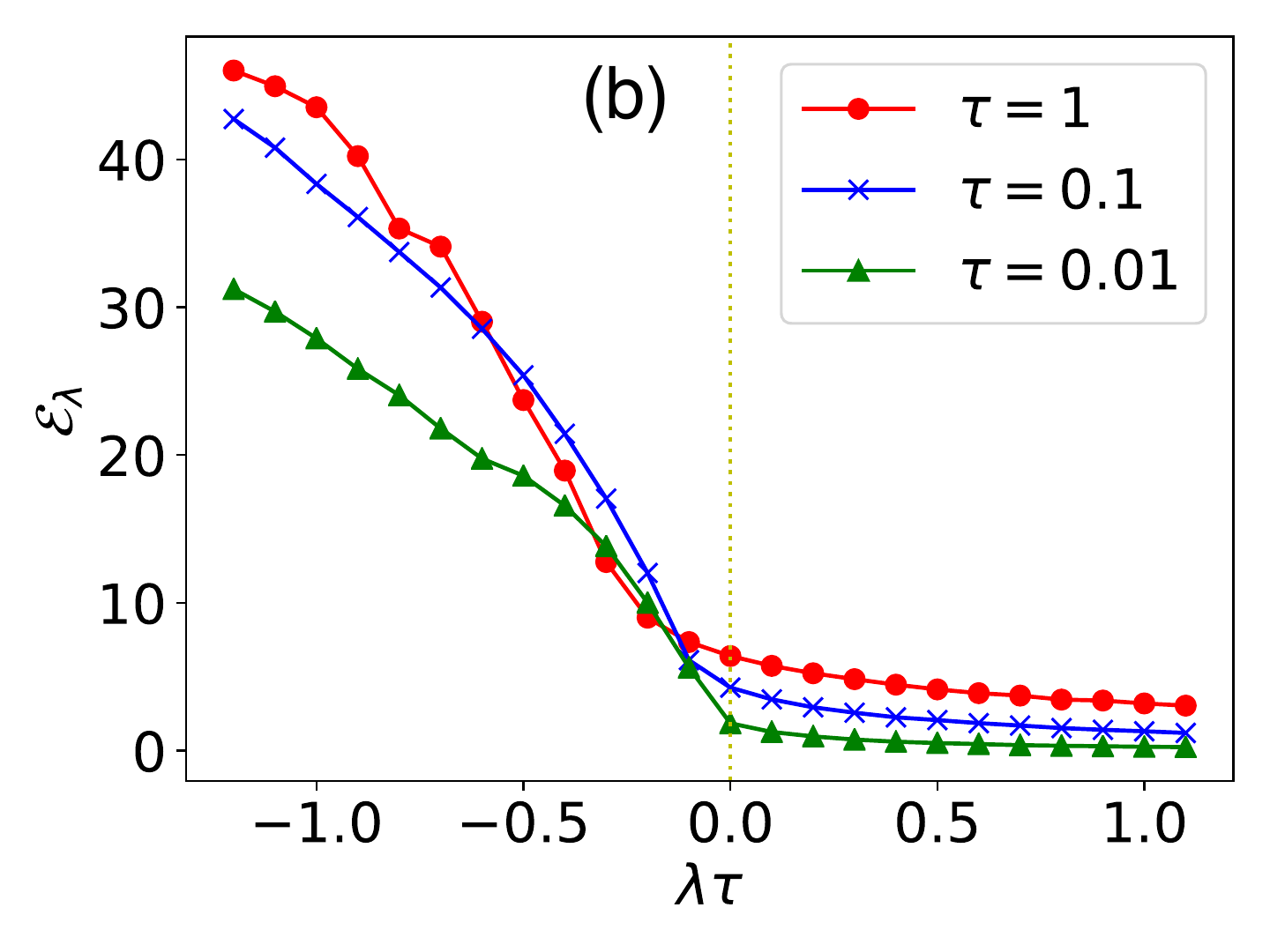}
	\caption{\label{fig:psi}
		(a)~Scaled rate function $\tau[\lambda\,{\cal E}_\lambda + \psi(\lambda)]$ and (b)~efficiency in the biased ensemble ${\cal E}_{\lambda}=\tau \langle (\nabla_jU)^2 \rangle_\lambda / (d N T_\text{\tiny A})$ as functions of the scaled bias parameter $\lambda \tau$ for different values of the persistence time $\tau$. The efficiency decreases rapidly for negative $\lambda\tau$ and vanishes at large positive $\lambda\tau$.	Simulations are performed using the population dynamics algorithm~\cite{Kurchan2006,Nemoto2016} (see Appendix A of~\cite{Nemoto2016} for the details of the algorithm), with number of clones given by $N_\text{c}=6400$.
	Parameter values: $L=10$, $a=1$, $\rho_0=0.32$, $V_\text{\tiny M}=10^{-1}$, $T=0$, $T_\text{\tiny A}=1$.
	}
\end{figure}

In that respect, the auxiliary dynamics~\eqref{eq:dyn_aux} can be regarded as analogous to a quorum-sensing dynamics where the self-propulsion adapts to local properties of the system~\cite {Tailleur2008, Solon2015}. In practice, large clusters should emerge spontaneously whenever the amplitude of self-propulsion fluctuations are reduced in dense regions, potentially up to a complete phase separation~\cite{Tailleur2008, Cates2015}. From the reasoning presented above, this condition is satisfied for $\kappa<0$. According to~\eqref{eq:kappa_aux}, the sign of $\kappa$ is determined by $\lambda\tau(\nabla_jU)^2/(dNT_\text{\tiny A})+\psi(\lambda)$, which we approximate by $\lambda\,{\cal E}_\lambda+\psi(\lambda)$ using self-averaging property, where ${\cal E}_\lambda=\tau\langle(\nabla_jU)^2\rangle_\lambda/(dNT_\text{\tiny A})$ is the efficiency in the biased ensemble.

To evaluate numerically $\lambda\,{\cal E}_\lambda+\psi(\lambda)$ as a function of $\lambda$, we sample the biased ensemble of the original dynamics~\eqref{eq:dyn}, where we take $T=0$ for convenience. The sampling is done using a population dynamics algorithm as detailed in~\cite{Kurchan2006, Nemoto2016}. We consider purely repulsive particles in two dimensions, with short-range pair-wise interactions given by the WCA potential $V({\bf r}) = V_\text{\tiny M} [(a/r)^{12}-2(a/r)^6] \Theta(a-r)$\footnote{This definition is equivalent to $V({\bf r}) = 4 V_\text{\tiny M} [(\sigma/r)^{12}-(\sigma/r)^6] \Theta( 2^{1/6} \sigma -r)$ when we set $\sigma = 2^{-1/6} a $.}. Our measurements show that $\lambda\,{\cal E}_\lambda+\psi(\lambda)$ is negative for all $\lambda$, see Fig.~\ref{fig:psi}(a), and decreases rapidly for $\lambda<0$. Then, we deduce that cluster formation should be strongly enhanced when $\lambda<0$: this is consistent with the efficiency ${\cal E}_\lambda$ taking large values in this regime, as shown in Fig.~\ref{fig:psi}(b). Note that ${\cal E}_\lambda$ is no longer bounded as in the unbiased dynamics~\eqref{eq:dyn}, it can now take values beyond $1$.

\begin{figure}
	\centering
	\includegraphics[width=.49\columnwidth]{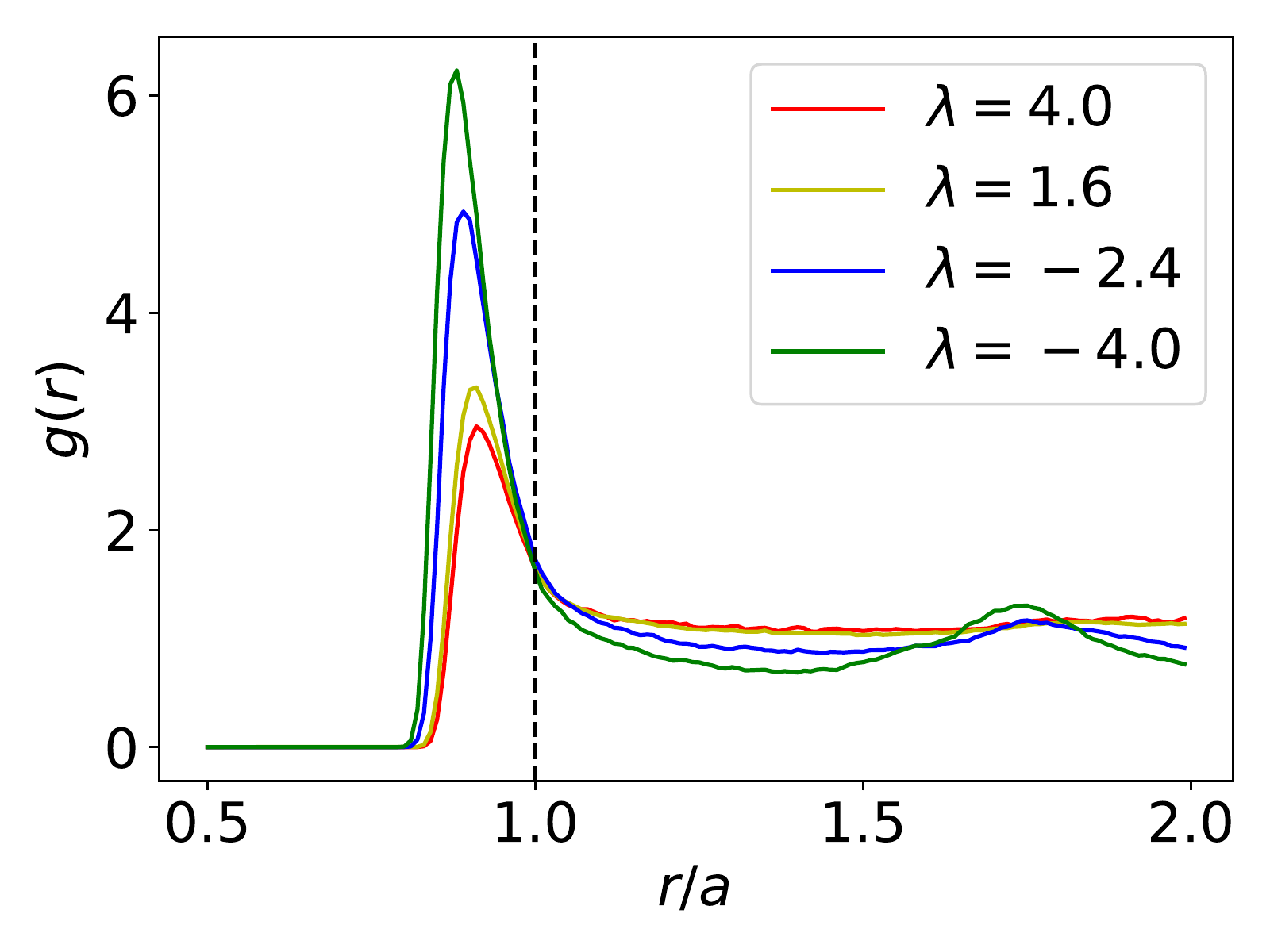}
	\caption{\label{fig:struct}
		Density pair correlation in the biased ensemble $g$ as a function of the scaled inter-particle distance $r/a$ for different values of the bias parameter $\lambda$. Numerical results show that biasing towards high and low efficiency leads respectively to increase and reduce the peaks of pair correlation for $\lambda<0$ and $\lambda>0$.
		Parameter values: $L=7$, $a=1$, $\rho_0=0.33$, $V_\text{\tiny M}=10^{-1}$, $\tau=10^{-1}$, $T=0$, $T_\text{\tiny A}=1$.
	}
\end{figure}

\begin{figure}
	\centering
	\includegraphics[width=\columnwidth]{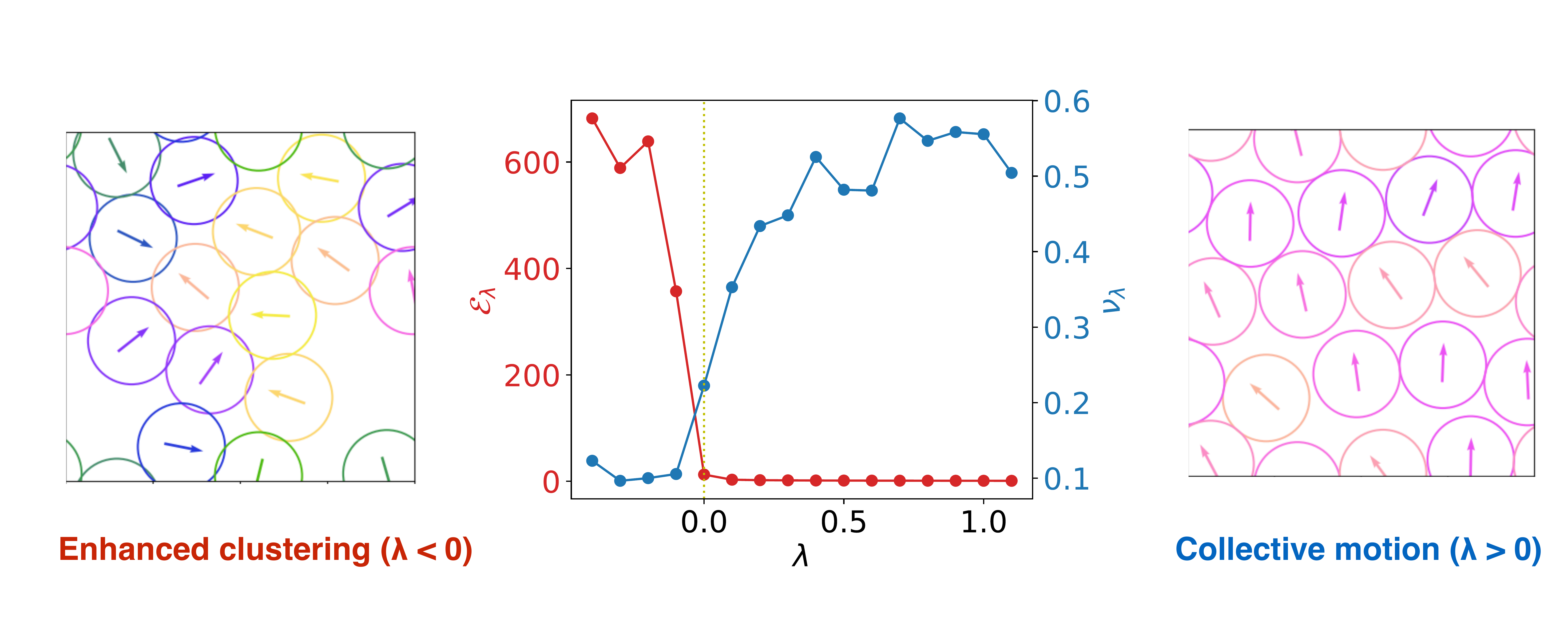}
	\caption{\label{fig:snap}
		Efficiency ${\cal E}_{\lambda}=\tau \langle (\nabla_jU)^2 \rangle_\lambda / (d N T_\text{\tiny A})$ (red) and collective orientation $\nu_\lambda = \langle |\sum_i {\bf v}_i| / |{\bf v}_i| \rangle_\lambda / N $ (blue) in the biased ensemble as functions of the bias parameter $\lambda$. Snapshots on the left and right, where colors refer to particle orientations, show respectively enhanced clustering at high efficiency ($\lambda=-0.3$), and collective directed motion at low efficiency ($\lambda=10$). This illustrates how biasing trajectories according to efficiency affects the collective dynamics.
		Parameter values: $L=4$, $a=1$, $\rho_0=1$ $V_\text{\tiny M}=10^{-1}$, $\tau=10^2$, $T=0$, $T_\text{\tiny A}=1$. Movies in~\cite{Supplemental}.
	}
\end{figure}

To determine quantitatively how the bias affects the emerging structure, we measure the pair correlation function of density $g$ in the biased ensemble, as shown in Fig.~\ref{fig:struct}. For positive and negative $\lambda$, the peaks of density correlations are respectively reduced and increased, thus showing that an effective attraction among particles is either tuned down or up when enforcing respectively low or high efficiency. In practice, such attractive effects lead to a complete phase separation at sufficiently high efficiency, as shown in Fig.~\ref{fig:snap}. This is analogous to the dynamical arrest already reported in diffusive passive systems~\cite{Lecomte2012, Sollich2015}.

Moreover, a polar phase without any equilibrium equivalent emerges at very low efficiency when the density and the persistent time are sufficiently large, as reported in Fig.~\ref{fig:snap}. A similar transition to collective directed motion was also obtained recently in~\cite{Nemoto2018} for active Brownian particles (ABPs) when biasing towards high dissipation, analogue to the low efficiency regime. At variance with ABPs, the amplitude of the self-propulsion is not fixed in our case, which allows the collective polarization to change direction more rapidly, see movie in~\cite{Supplemental}. We defer further investigations on the finite-size scaling of this transition to future works. Overall, our results illustrate how changing efficiency with a bias on trajectories affects dramatically the collective dynamics, thus providing a route to promote either phase separation or collective directed motion. We defer a quantitative study of finite size effects appearing in theses transitions to future works.

\section{Conclusion}

In this paper, we have explored how the local dissipation controls the emerging properties of active fluids, by investigating how it connects to the transport of particles and how it can trigger phase transitions. By mapping the random self-propulsion into a disordered drive, as introduced in~\cite{Suri2019}, we have put forward generic relations between the {\it collisional efficiency}, quantifying the tendency of collisions to stabilize clusters, and the displacement of a tracer immersed in the fluid. This approach could also be used to explore the connection between transport and dissipation in other active systems, for instance in aligning~\cite{Baskaran2008, Farrell2012, Chate2015} or self-spinning particles~\cite{Nguyen2014, Bartolo2016, Klymko2017}. In particular, it would be interesting to investigate the coupling between translational and orientational dynamics of spinners, for which atypical transverse response has been reported recently~\cite{Vitelli2017, Vitelli2018}.

Using tools of large deviations~\cite{Lecomte2007, Touchette2009, Jack2010}, we have also shown that tuning efficiency with a specific control parameter modifies the emergent dynamics: one can promote either a phase separation or a collective directed motion. We have proposed an auxiliary dynamics which captures explicitly the effect of the control parameter on interactions, thus providing a microscopic mechanism for phase separation. Overall, these results suggest that phase diagrams of active fluids, usually determined by P\'eclet number and density~\cite{Redner2013, Solon2018, Junco2018a}, can now also be described in terms of a parameter conjugate to energy transfers.

It would be interesting to bias active dynamics with other nonequilibrium observables, for instance the entropy production quantifying the irreversibility of the dynamics~\cite{Speck2016, Nardini2016, Mandal2017, Seifert2018, Shankar2018, Bo2019}. More generally, one could explore the effect of dynamical bias in other active systems, for instance featuring a flocking transition~\cite{Vicsek1995, Toner1995} or a rigidity-induced phase transition~\cite{Manning2015, Manning2016, Marchetti2017}, either from a particle-based or a hydrodynamic perspective. In short, our work opens the door to the search of new phases and dynamics, as well as unexpected transitions between them, in biased ensembles of active matter.


\section*{Acknowledgements}
	The authors acknowledge insightful discussions with Michael E. Cates, Vincent D\'emery, Robert L. Jack, and Julien Tailleur. This work was granted access to the HPC resources of CINES/TGCC under the allocation 2018-A0042A10457 made by GENCI and of MesoPSL financed by the Region Ile de France and the project Equip@Meso (reference ANR-10-EQPX-29-01) of the program Investissements d'Avenir supervised by the Agence Nationale pour la Recherche. SV acknowledges support from the Sloan Foundation, startup funds from the University of Chicago and support from the National Science Foundation under award number DMR-1848306. \'EF benefits from an Oppenheimer Research Fellowship from the University of Cambridge, and a Junior Research Fellowship from St Catherine's College.


\appendix

\section{Efficiency and mobility}\label{app:tracer}

In this appendix, we derive an expression for efficiency $\cal E$ and linear mobility $\mu$ in terms of microscopic details. To this aim, we employ a perturbative treatment for weak interactions between a given tagged tracer and the surrounding particles. This amounts to using the bath-tracer interaction parameter $h$ in the effective tracer dynamics~\eqref{eq:dyn_tracer} as an expansion parameter.

The probability for the tracer to follow a given trajectory during the time interval $[0,t]$ is determined by the path weight $\mathbb{P} [\{{\bf r}_i\}_0^t] \sim \ee^{-{\cal A}[\{{\bf r}_i\}_0^t]} $, where $\cal A$ refers to the dynamic action. It can be split into three contributions as ${\cal A}={\cal A}_0+h{\cal A}_\text{d}+h^2{\cal A}_\text{int}$, which respectively concern the non-interacting part, the driving part and the interacting part of the dynamics. Using standard path integral techniques~\cite{Martin1973, Dominicis1975}, we write these contributions explicitly from the tracer dynamics~\eqref{eq:dyn_tracer} in terms of the tracer position ${\bf r}_0$ and the conjugated process $\bar{\bf r}_0$ as
\begin{equation}\label{eq:action}
	\begin{aligned}
		{\cal A}_0 &= \int \bar{\bf r}_0(s) \cdot \big[ \ii ( \dot{\bf r}_0(s) - \tilde{\bf f}_0(s) ) + (T+\sigma T_\text{\tiny A}) \bar{\bf r}_0(s) \big] \dd s ,
		\\
		{\cal A}_\text{d} &= \int\dd s\dd u\int_{\bf k} {\cal G}_{\bf k}(s-u) \bar V_{\bf k} \ee^{\ii{\bf k}\cdot{\bf r}_0(s)} \big[\ii {\bf k}\cdot\bar{\bf r}_0 (s)\big] \big[ {\bf k}\cdot\bar{\bf F}_{\bf k}(u) \big] ,
		\\
		{\cal A}_\text{int} &= \rho_0 \int\dd s\dd u\int_{\bf k} (|{\bf k}|\bar V_{\bf k})^2 \ee^{\ii{\bf k}\cdot[{\bf r}_0(s)-{\bf r}_0(u)]} {\cal G}_{\bf k}(s-u) \bar{\bf r}_0 (s) \cdot \bigg[ \frac{(T+\sigma T_\text{\tiny A}) \bar{\bf r}_0(u)}{T+\sigma T_\text{\tiny A}+\rho_0V_{\bf k}} - {\bf k} \bigg] .
	\end{aligned}
\end{equation}
The contributions from the drive ${\cal A}_\text{d}$ and from the interactions ${\cal A}_\text{int}$ gather all the non-linearities in the tracer dynamics.

To evaluate the tracer velocity appearing in dissipation ${\cal J} = \langle{\bf f}_i\cdot\langle\dot{\bf r}_i\rangle\rangle_\text{d}$ and mobility $\mu = \underset{t\to\infty}{\lim} \langle\langle\dot{\bf r}_0(t)\rangle\rangle_\text{d}\cdot\hat{\bf e}/ f_\text{\tiny P}$, where $\langle\cdot\rangle$ denotes here an average at fixed disorder, we expand the path probability $\mathbb{P}$ in terms of the small parameter $h$ for weak interactions. Following~\cite{Demery2011, Demery2014}, this lets us with expectation values with respect to only the non-interacting part of the action ${\cal A}_0$, to be evaluated as Gaussian integrals. The tracer velocity then reads
\begin{equation}\label{eq:pert}
	\langle\dot{\bf r}_0\rangle = \langle\dot{\bf r}_0\rangle_0 - h \langle{\cal A}_\text{d}\dot{\bf r}_0\rangle_0 + h^2 \langle({\cal A}_\text{d}^2/2-{\cal A}_\text{int})\dot{\bf r}_0\rangle_0 + {\cal O}(h^3) ,
\end{equation}
where $\langle\cdot\rangle_0$ denotes an average with respect to ${\cal A}_0$. The leading order is simply deduced as $\langle\dot{\bf r}_0\rangle_0 = {\bf f}_0$, yielding $ {\cal J} = \langle{\bf f}_i^2\rangle_\text{d} = d N T_\text{\tiny A}/\tau $ in the absence of interactions, as expected. Substituting the explicit expression of ${\cal A}_\text{int}$ and ${\cal A}_\text{d}$ from~\eqref{eq:action} in~\eqref{eq:pert}, the corrections up to order $h^2$ read
\begin{equation}\label{eq:action_g}
	\begin{aligned}
		\langle{\cal A}_\text{d}\dot{\bf r}_0\rangle_0 &= \int\dd s\dd u\int_{\bf k} {\cal G}_{\bf k}(s-u) \bar V_{\bf k} \Big\langle\dot{\bf r}_0(t) \big[\ii {\bf k}\cdot\bar{\bf r}_0 (s)\big] \ee^{\ii{\bf k}\cdot{\bf r}_0(s)} \Big\rangle_0 \big[ {\bf k}\cdot\bar{\bf F}_{\bf k}(u) \big] ,
		\\
		\langle{\cal A}_\text{int}\dot{\bf r}_0\rangle_0 &= \rho_0 \int\dd s\dd u\int_{\bf k} (|{\bf k}|\bar V_{\bf k})^2 {\cal G}_{\bf k}(s-u) \bigg\{ - \Big\langle \dot{\bf r}_0(t) \big[{\bf k}\cdot\bar{\bf r}_0 (s)\big] \ee^{\ii{\bf k}\cdot[{\bf r}_0(s)-{\bf r}_0(u)]} \Big\rangle_0
		\\
		&\qquad\qquad\qquad + \frac{T+\sigma T_\text{\tiny A}}{T+\sigma T_\text{\tiny A}+\rho_0V_{\bf k}} \Big\langle \dot{\bf r}_0(t) \big[\bar{\bf r}_0 (s)\cdot\bar{\bf r}_0(u)\big] \ee^{\ii{\bf k}\cdot[{\bf r}_0(s)-{\bf r}_0(u)]} \Big\rangle_0 \bigg\} ,
	\\
		\langle{\cal A}_\text{d}^2\dot{\bf r}_0\rangle_0 &= - \int\dd s\dd s'\dd u\dd u'\int_{{\bf k},{\bf k}'} {\cal G}_{\bf k}(s-u) {\cal G}_{{\bf k}'}(s'-u') \bar V_{\bf k}\bar V_{{\bf k}'} \big[ {\bf k}\cdot\bar{\bf F}_{\bf k}(u) \big] \big[ {\bf k}'\cdot\bar{\bf F}_{{\bf k}'}(u') \big]
		\\
		&\qquad\qquad\qquad \times \Big\langle\dot{\bf r}_0(t) \big[{\bf k}\cdot\bar{\bf r}_0 (s)\big] \big[{\bf k}'\cdot\bar{\bf r}_0 (s')\big] \ee^{\ii{\bf k}\cdot{\bf r}_0(s)+\ii{\bf k}'\cdot{\bf r}_0(s')} \Big\rangle_0 ,
	\end{aligned}
\end{equation}
where ${\cal G}_{\bf k}(t) = \ee^{-{\bf k}^2(T+\sigma T_\text{\tiny A}+\rho_0 V_{\bf k})t} \Theta(t) $. The first term $\langle{\cal A}_\text{d}\dot{\bf r}_0\rangle_0$ vanishes when averaging over disorder, since $\bar{\bf F}({\bf r},t) = \sum_{i\neq0}\tilde{\bf f}_i(t)\delta[{\bf r}-{\bf r}_i(t)]$ has zero average and it is uncorrelated with $\tilde{\bf f}_0$, and we assume that $\langle{\cal A}_\text{d}\dot{\bf r}_0\rangle_0$ also vanishes when averaging over disorder, which amounts to setting the two-point correlations of $\bar{\bf F}$ to zero. Thus, the effect of the bath on the tracer dynamics is effectively analogue to that of passive Brownian particles at temperature $T+\sigma T_\text{\tiny A}$.

Following~\cite{Demery2011, Demery2014}, we then use Wick's theorem for exponential observables to evaluate the correlation functions between the tracer position ${\bf r}_0$ and the conjugate variable $\bar{\bf r}_0$ as
\begin{equation}\label{eq:corr}
	\begin{aligned}
		\Big\langle \dot{\bf r}_0(t) \big[{\bf k}\cdot\bar{\bf r}_0(s)\big] \ee^{\ii{\bf k}\cdot[{\bf r}_0(u)-{\bf r}_0(s)]} \Big\rangle_0 &= \ii{\bf k}\delta(t-s)\ee^{-{\bf k}^2(T+\sigma T_\text{\tiny A})(t-u)}
		\\
		&\quad \times \ee^{\ii{\bf k}\cdot\big[ f_\text{\tiny P} \hat{\bf e} (t-u) + \int_u^t \tilde{\bf f}_0(w) \dd w \big]} ,
		\\
		\Big\langle \dot{\bf r}_0(t) \big[\bar{\bf r}_0(s)\cdot\bar{\bf r}_0(u)\big] \ee^{\ii{\bf k}\cdot[{\bf r}_0(u)-{\bf r}_0(s)]} \Big\rangle_0 &= -\ii {\bf k} \delta(t-s)\ee^{-{\bf k}^2(T+\sigma T_\text{\tiny A})(t-u)}
		\\
		&\quad \times \ee^{\ii{\bf k}\cdot\big[ f_\text{\tiny P} \hat{\bf e} (t-u) + \int_u^t \tilde{\bf f}_0(w) \dd w \big]} ,
	\end{aligned}
\end{equation}
where $f_\text{\tiny P} \hat{\bf e}$ is the constant perturbation force. Averaging over disorder then requires to evaluate the following correlation functions:
\begin{equation}\label{eq:corr_dis}
	\begin{aligned}
		\Big\langle \ee^{-\ii{\bf k}\cdot\int_u^t \tilde{\bf f}_0(w) \dd w} \Big\rangle_\text{d} &= \ee^{ {\bf k}^2 T_\text{\tiny A} \big[ \sigma (t-u) - \int \frac{\phi(\omega)}{\tau} \frac{1-\ee^{\ii\omega (t-u)}}{\omega^2} \frac{\dd\omega}{2\pi} \big] } ,
		\\
		-\ii{\bf k}\cdot\Big\langle{\bf f}_0(t) \ee^{-\ii{\bf k}\cdot\int_u^t \tilde{\bf f}_0(w) \dd w} \Big\rangle_\text{d} &= \ee^{ {\bf k}^2 T_\text{\tiny A} \sigma (t-u) } \frac{\dd}{\dd t} \big\langle \ee^{-\ii{\bf k}\cdot\int_u^t {\bf f}_0(w) \dd w} \big\rangle_\text{d}  
		\\
		&= \ii {\bf k}^2 T_\text{\tiny A} \ee^{ {\bf k}^2 T_\text{\tiny A} \big[ \sigma (t-u) - \int \frac{\phi(\omega)}{\tau} \frac{1-\ee^{\ii\omega (t-u)}}{\omega^2} \frac{\dd\omega}{2\pi} \big] } \int \frac{\phi(\omega')}{\tau} \frac{\ee^{\ii\omega'(t-u)}}{\omega'} \frac{\dd\omega'}{2\pi} ,
	\end{aligned}
\end{equation}
where we have used~\eqref{eq:drive_corr} and~\eqref{eq:drive_corr_g}. The corrections to order $h^2$ in $\langle{\bf f}_0\cdot\langle\dot{\bf r}_0\rangle\rangle_\text{d}$ and $\langle\langle\dot{\bf r}_0(t)\rangle\rangle_\text{d}\cdot\hat{\bf e}$ then read
\begin{equation}
	\begin{aligned}
		\langle {\bf f}_0 \cdot \langle {\cal A}_\text{int} \dot{\bf r}_0 \rangle_0 \rangle_\text{d} \big|_{f_\text{\tiny P}=0} &= \frac{\rho_0 T_\text{\tiny A}}{d} \int_{\bf k} ({\bf k}^2V_{\bf k})^2 \frac{2(T+\sigma T_\text{\tiny A})+\rho_0V_{\bf k}}{T+\sigma T_\text{\tiny A}+\rho_0V_{\bf k}}
		\\
		&\quad\times \int_0^\infty \dd t \ee^{ - {\bf k}^2 \big[ (2T+\sigma T_\text{\tiny A}+\rho_0V_{\bf k}) t + T_\text{\tiny A} \int \frac{\phi(\omega)}{\tau} \frac{1-\ee^{\ii\omega t}}{\omega^2} \frac{\dd\omega}{2\pi} \big] } \int \frac{\phi(\omega')}{\tau} \frac{\ee^{\ii\omega't}}{\ii\omega'} \frac{\dd\omega'}{2\pi} ,
		\\
		\langle \langle {\cal A}_\text{int} \dot{\bf r}_0 \rangle_0 \rangle_\text{d} \cdot \hat{\bf e} &= \frac{f_\text{\tiny P}\rho_0}{d} \int_{\bf k} ({\bf k}^2V_{\bf k})^2 \frac{2(T+\sigma T_\text{\tiny A})+\rho_0V_{\bf k}}{T+\sigma T_\text{\tiny A}+\rho_0V_{\bf k}}
		\\
		&\quad \times \int_0^\infty t \dd t \ee^{ - {\bf k}^2  \big[ (2T+\sigma T_\text{\tiny A}+\rho_0V_{\bf k}) t + T_\text{\tiny A} \int \frac{\phi(\omega)}{\tau} \frac{1-\ee^{\ii\omega t}}{\omega^2} \frac{\dd\omega}{2\pi} \big] } + {\cal O}(f_\text{\tiny P}^2) .
	\end{aligned}
\end{equation}
For the distribution~\eqref{eq:g}, we get
\begin{equation}
	\begin{aligned}
		\int \frac{\phi(\omega)}{\tau} \frac{1-\ee^{\ii\omega t}}{\omega^2} \frac{\dd\omega}{2\pi} &= t + \tau ( \ee^{-t/\tau} - 1 ) ,
		\\
		\int \frac{\phi(\omega)}{\tau} \frac{\ee^{\ii\omega t}}{\ii\omega} \frac{\dd\omega}{2\pi} &= \ee^{-t/\tau} - 1 .
	\end{aligned}
\end{equation}
Finally, the leading contribution to efficiency ${\cal E}=1-\tau{\cal J}/dNT_\text{\tiny A}$ and reduced mobility $1-\mu$ then follows as
\begin{equation}
	\begin{aligned}\label{eq:eff}
		{\cal E} &= \frac{h^2\rho_0\tau}{d} \int_{\bf k} ({\bf k}^2V_{\bf k})^2 \frac{2(T+\sigma T_\text{\tiny A})+\rho_0V_{\bf k}}{T+\sigma T_\text{\tiny A}+\rho_0V_{\bf k}}
		\\
		&\qquad \times \int_0^\infty \dd t (1-\ee^{-t/\tau}) \ee^{ - {\bf k}^2 \big[ (2T+(\sigma+1) T_\text{\tiny A}+\rho_0V_{\bf k}) t + T_\text{\tiny A} \tau ( \ee^{-t/\tau} - 1 ) \big] } + {\cal O}(h^3) ,
	\end{aligned}
\end{equation}
and
\begin{equation}\label{eq:mob}
	\begin{aligned}
		1-\mu &= \frac{h^2\rho_0}{d} \int_{\bf k} ({\bf k}^2V_{\bf k})^2 \frac{2(T+\sigma T_\text{\tiny A})+\rho_0V_{\bf k}}{T+\sigma T_\text{\tiny A}+\rho_0V_{\bf k}}
	\\
	&\qquad\times \int_0^\infty t \dd t \ee^{ - {\bf k}^2  \big[ (2T+(\sigma+1) T_\text{\tiny A}+\rho_0V_{\bf k}) t + T_\text{\tiny A}\tau ( \ee^{-t/\tau} - 1 ) \big] } + {\cal O}(h^3) .
	\end{aligned}
\end{equation}
The explicit integration over the modes is done with the Fourier transform of the potential $V({\bf r}) = V_\text{\tiny M}(1-r/a)^2 \Theta(a-r)$, given in two dimensions by $V_{\bf k} = (2\pi V_\text{\tiny M}/k^2)\big\{ \pi\big[J_1(ak)H_0(ak)-J_0(ak)H_1(ak)\big] - 2 J_2(ak)\big\}$, where $J_n$ and $H_n$ respectively denote the Bessel and Struve functions of order $n$.


\section{Entropy production rate of coarse-grained dynamics}\label{app:ent}

This appendix is devoted to deriving the entropy production rate $\Sigma$ associated with the coarse-grained dynamics~\eqref{eq:dyn_dens} at fixed disorder. We introduce an effective
free-energy $\cal F$ as
\begin{equation}
	{\cal F} = \int \rho \big[ (T+\sigma T_\text{\tiny A}) (\ln\rho-1) + U \big] \dd{\bf r} ,
\end{equation}
so that the fluctuating hydrodynamic equation~\eqref{eq:dyn_dens} can be written as
\begin{equation}
	\partial_t\rho = \nabla\cdot\bigg[ \rho \nabla\frac{\delta{\cal F}}{\delta\rho} - {\bf F} + \sqrt{2\rho(T+\sigma T_\text{\tiny A})}{\boldsymbol\Lambda} \bigg] .
\end{equation}
Following~\cite{Leonard2013}, the corresponding dynamic action ${\cal A}_\rho$, defined from the probability weight as $\mathbb{P}_\rho[\{\rho\}_0^t] \sim \ee^{-{\cal A}_\rho[\{\rho\}_0^t]} $, reads
\begin{equation}\label{eq:action_rho}
	\begin{aligned}
		{\cal A}_\rho = &\frac{1}{4(T+\sigma T_\text{\tiny A})} \int\dd{\bf r}\dd{\bf r}'\dd t \bigg\{\partial_t \rho - \nabla\cdot\bigg[ \rho \nabla\frac{\delta{\cal F}}{\delta\rho} - {\bf F} \bigg]\bigg\}\bigg|_{\bf r}
		\\
		&\quad\quad\quad\quad\quad \times {\cal K}({\bf r},{\bf r}') \bigg\{\partial_t \rho - \nabla\cdot\bigg[ \rho \nabla\frac{\delta{\cal F}}{\delta\rho} - {\bf F} \bigg]\bigg\}\bigg|_{{\bf r}'} .
	\end{aligned}
\end{equation}
The operator $\cal K$ satisfies
\begin{equation}
	\int \dd{\bf r}'' {\cal K}({\bf r},{\bf r}'') \nabla_{{\bf r}'} \nabla_{{\bf r}''} \big[ \rho({\bf r}') \delta({\bf r}'-{\bf r}'') \big] = \delta({\bf r}-{\bf r}') ,
\end{equation}
whose explicit solution can be written as
\begin{equation}\label{eq:kappa}
	\nabla_{\bf r} {\cal K}({\bf r},{\bf r}') = - \frac{1}{\rho({\bf r})} \nabla_{\bf r} G({\bf r}-{\bf r}') ,
\end{equation}
where $G$ is the Green function of the Laplace operator:
\begin{equation}\label{eq:G}
	\nabla_{\bf r}^2 G({\bf r}-{\bf r}') = \delta({\bf r}-{\bf r}') .
\end{equation}
The action associated with the backward dynamics is deduced from the bare action~\eqref{eq:action_rho} under the following time-reversal transformation
\begin{equation}
	\begin{aligned}
		\rho({\bf r},t)&\longrightarrow\rho({\bf r},-t),
		\\
		\partial_t\rho({\bf r}, t)&\longrightarrow-\partial_t\rho({\bf r}, -t),
		\\
		{\bf F}({\bf r},t)&\longrightarrow{\bf F}({\bf r},-t) .
	\end{aligned}
\end{equation}
The entropy production rate $\Sigma$, defined in~\eqref{eq:sigma}, then follows as
\begin{equation}
	\Sigma = \int\frac{\dd{\bf r}\dd{\bf r}'}{T+\sigma T_\text{\tiny A}}\bigg\langle\partial_t \rho({\bf r}) {\cal K}({\bf r},{\bf r}') \nabla\cdot\bigg[ \rho \nabla\frac{\delta{\cal F}}{\delta\rho} - {\bf F} \bigg]\bigg|_{{\bf r}'} \bigg\rangle ,
\end{equation}
where we have replaced a time average by an ensemble average using ergodicity. From~(\ref{eq:kappa}-\ref{eq:G}), $\Sigma$ can then be simplified by successive integration by parts, yielding
\begin{equation}
	\Sigma = \frac{1}{T+\sigma T_\text{\tiny A}} \int \dd{\bf r} \bigg[- \bigg\langle\frac{\delta{\cal F}}{\delta\rho}\partial_t\rho \bigg\rangle + \bigg\langle\frac{{\bf V}\cdot{\bf F}}{\rho}\bigg\rangle \bigg] ,
\end{equation}
where we have used $\partial_t\rho=-\nabla\cdot{\bf V}$. The first term vanishes in steady state, so that the entropy production rate reduces to the final expression given in~\eqref{eq:sigma_g}.


\section{Auxiliary dynamics}\label{app:bias}

In this Appendix, we derive the auxiliary dynamics associated with biasing with the time-extensive efficiency $\varepsilon$ in~\eqref{eq:eff_sc}, whose operator ${\cal L}_\text{aux}$ is defined in terms of $\Phi$ in~\eqref{eq:aux}. To this aim, we obtain explicitly $\Phi$ from the eigenvalue problem in~\eqref{eq:eigen} by using a perturbative treatment at small persistence time $\tau$. The Fokker-Planck operator ${\cal L}$ associated with the dynamics~\eqref{eq:dyn_p} reads
\begin{equation}\label{eq:L}
	{\cal L} = -{\bf p}_i\cdot\nabla_i + \frac{1}{\tau} \frac{\partial}{\partial{\bf p}_i}\cdot\bigg[{\bf p}_i + (1+\tau{\bf p}_j\cdot\nabla_j)\nabla_iU + \frac{T_\text{\tiny A}}{\tau} \frac{\partial}{\partial{\bf p}_i} \bigg] .
\end{equation}
Following previous works~\cite{Nardini2016}, we scale the particle velocity as $\tilde{\bf p}_i=\sqrt\tau{\bf p}_i$, so that the scaled version $\tilde{\cal L}$ of ${\cal L}$ is given by
\begin{equation}
	\tilde{\cal L} = -\frac{\tilde{\bf p}_i\cdot\nabla_i}{\sqrt\tau} + \frac{1}{\tau} \frac{\partial}{\partial\tilde{\bf p}_i}\cdot\bigg[\tilde{\bf p}_i + (\sqrt\tau+\tau\tilde{\bf p}_j\cdot\nabla_j)\nabla_iU + T_\text{\tiny A} \frac{\partial}{\partial\tilde{\bf p}_i} \bigg] .
\end{equation}
The stationary statistics of position and velocity is then given by the equilibrium-like Maxwell-Boltzmann distribution $\ee^{-[U(\{{\bf r}_i\})+\tilde{\bf p}_i^2/2]/T_\text{\tiny A}}$ at leading order in $\tau$. Moreover, we scale the conjugate parameter as $\tilde\lambda=\lambda\tau$. Replacing $\cal L$ by $\tilde{\cal L}$ in~\eqref{eq:eigen} and expanding $\Phi = \tau \tilde\Phi + {\cal O}(\tau^{3/2}) $, the leading order of the eigenvalue problem reads
\begin{equation}\label{eq:eigen_b}
	\bigg[T_\text{\tiny A}\frac{\partial}{\partial\tilde{\bf p}_i} -\tilde{\bf p}_i \bigg] \cdot \frac{\partial\tilde\Phi}{\partial\tilde{\bf p}_i} = \frac{\tilde\lambda}{dNT_\text{\tiny A}}(\nabla_iU)^2 + \psi(\tilde\lambda) .
\end{equation}
Given that there is no coupling between particle velocity and position at this order, we deduce that $\tilde\Phi$ should be a function of $\tilde p=\sqrt{\sum_i\tilde{\bf p}_i^2}$ and $\{{\bf r}_i\}$ only. As a result, the lhs of~\eqref{eq:eigen_b} can be cast in the form
\begin{equation}
	\bigg[T_\text{\tiny A}\frac{\partial}{\partial\tilde{\bf p}_i} -\tilde{\bf p}_i \bigg] \cdot \frac{\partial\tilde\Phi}{\partial\tilde{\bf p}_i} = \bigg[T_\text{\tiny A} \bigg(\frac{\partial}{\partial\tilde p} + \frac{dN-1}{\tilde p}\bigg) - \tilde p\bigg] \frac{\partial\tilde\Phi}{\partial\tilde p} .
\end{equation}
The explicit solution for $\partial\tilde\Phi/\partial\tilde p$ then follows as
\begin{equation}\label{eq:phi}
	\frac{\partial\tilde\Phi}{\partial\tilde p} = - \bigg[ \frac{\tilde\lambda}{dNT_\text{\tiny A}}(\nabla_jU)^2 + \psi(\tilde\lambda) \bigg] \frac{\tilde p\,\ee^{\tilde p^2/(2T_\text{\tiny A})}}{2T_\text{\tiny A}} E_{1-dN/2}\bigg(\frac{\tilde p^2}{2T_\text{\tiny A}}\bigg) ,
\end{equation}
where $E_n(z) = \int_1^\infty \dd t \ee^{-zt}/t^n$ is the exponential integral function. At next order, the eigenvalue problem involves a coupling between position and velocity, so that the corresponding differential equation can no longer be simplified. This leads us to restrict the derivation of the auxiliary dynamics to first order. Substituting this result into the auxiliary operator~\eqref{eq:aux}, the auxiliary dynamics follows in bare units as
\begin{equation}\label{eq:dyn_aux_b}
	\tau\dot{\bf p}_i = - \bigg\{ \bigg[ \frac{\lambda\tau}{dNT_\text{\tiny A}}(\nabla_kU)^2 + \psi(\lambda) \bigg] \ee^{\tau {\bf p}_j^2/(2T_\text{\tiny A})} E_{1-dN/2}\bigg(\frac{\tau {\bf p}_j^2}{2T_\text{\tiny A}}\bigg) + 1 \bigg\} {\bf p}_i - (1+\tau{\bf p}_j\cdot\nabla_j)\nabla_iU + {\boldsymbol\eta}_i ,
\end{equation}
where we have used the chain rule $\partial\Phi/\partial{\bf p}_i=({\bf p}_i/p)\,\partial\Phi/\partial p$. To gain physical insight on the effect of the control parameter $\lambda$, the auxiliary dynamics~\eqref{eq:dyn_aux_b} can be cast in a form similar to the original dynamics~\eqref{eq:dyn} in terms of position and self-propulsion, as given in~\eqref{eq:dyn_aux}.


\section*{References}

\bibliographystyle{iopart-num}
\bibliography{Efficiency_ref}

\end{document}